\documentclass[10pt, prl, preprintnumbers, aps, twocolumn, floatfix, superscriptaddress,nofootinbib]{revtex4-2}

\usepackage[dvipsnames]{xcolor}
\usepackage[english]{babel}
\usepackage{caption}
\usepackage{subcaption}
\usepackage{relsize}
\usepackage{nicematrix}
\usepackage[letterpaper,top=2cm,bottom=2cm,left=2cm,right=2cm,marginparwidth=1.75cm]{geometry}

\usepackage{amsmath}
\usepackage{amssymb}
\usepackage{MnSymbol}
\usepackage{mathtools}
\usepackage{wasysym}
\usepackage{graphicx}
\usepackage{graphics}
\usepackage[colorlinks=true, allcolors=blue]{hyperref}
\usepackage{physics}
\usepackage{bm}

\usepackage{ragged2e} 
\usepackage{tikz}
\usepackage{feynmp-auto}
\usepackage[compat=1.1.0]{tikz-feynman}
\usetikzlibrary{bending}
\usetikzlibrary{shapes.geometric}
\usepackage[normalem]{ulem}

\newcommand{\slashed}[1]{#1\!\!\!/}

\newcommand{\threej}[6]{
    \begin{pmatrix}
        #1 & #3 & #5 \\
        #2 & #4 & #6
    \end{pmatrix}%
}
\newcommand{\CG}[6]{
    \bra{#1, #2, #3, #4}\ket{#5, #6}
}
\newcommand{\sixj}[6]{
    \left\{
    \begin{array}{ccc}
        #1 & #2 & #3 \\
        #4 & #5 & #6
    \end{array}
    \right\}
}
\allowdisplaybreaks

\begin{document}

\title{
Testing New Scalar Interactions\\ in Few-Electron Highly Charged Ions}
\date{\today}
\author{M.~Moretti}
\affiliation{Max-Planck-Institut für Kernphysik, Saupfercheckweg 1, 69117 Heidelberg, Germany}
\author{C.~de Jonge}
\affiliation{Max-Planck-Institut für Kernphysik, Saupfercheckweg 1, 69117 Heidelberg, Germany}
\author{J.~Jaeckel}
\affiliation{Institut f\"ur Theoretische Physik, Universit\"at Heidelberg,
Philosophenweg 16, 69120 Heidelberg, Germany}
\author{C.~H.~Keitel}
\affiliation{Max-Planck-Institut für Kernphysik, Saupfercheckweg 1, 69117 Heidelberg, Germany}
\author{Z.~Harman}
\affiliation{Max-Planck-Institut für Kernphysik, Saupfercheckweg 1, 69117 Heidelberg, Germany}

\begin{abstract}

We investigate how a hypothetical scalar boson mediating an interaction between electrons as well as between electrons and nucleons would affect the $g$ factor of lithium-like highly charged ions. In such ions, the strong nuclear Coulomb fields enhance electron–electron interactions, making them ideal systems for detecting subtle new physics signatures. Exceptionally accurate quantum electrodynamic predictions and experimental data in such few-electron systems allow for sensitive probes, thereby enabling bounds on the boson’s coupling strength.
Exploiting the enhanced sensitivity of highly charged ions to short-range interactions, we combine $g$-factor measurements and quantum electrodynamic theory predictions of lithium- and hydrogen-like ions with the free-electron magnetic moment and an isotope-shift measurement to constrain simultaneously the electron-proton, electron-neutron, and electron-electron coupling combinations as functions of the scalar mass.
We find that precision $g$-factor spectroscopy provides competitive constraints on scalar interactions over a broad mass range and, in particular, yields bounds on electron–electron interactions from bound-state QED observables.

\end{abstract}

\maketitle

\section{Introduction}

In recent years, there has been accelerating progress in testing extensions of the Standard Model (SM) at the low-energy precision frontier with atomic systems~\cite{JaeckelRingwald2010,Jaeckel2010,Delaunay2017,Flambaum2018,Berengut2018,Safronova2018,Jones2020,Counts2020,Rehbehn2021,Sailer2022,Rehbehn2023,Potvliege:2023lvf,Potvliege2025,Wilzewski2025,Door2025,Jaeckel:2026aeh}
 (see, e.g.~\cite{Ginges2004,Antypas:2022asj,Delaunay:2026ymq} for reviews).  This frontier provides a valuable complement to searches for physics beyond the SM, commonly referred to as New Physics (NP), at high energy accelerators, which explore the energy and intensity frontiers. 
At the precision frontier, rigorous searches for NP signatures can be performed with relatively compact experiments. Standard methods for testing quantum electrodynamics (QED), such as measurements of the Lamb shift, can also be extended to search for hypothetical NP effects.

In atomic physics, precise experimental measurements are often accompanied by advanced theoretical calculations.  High-precision calculations are particularly feasible for few-electron ions, making this an exceptionally powerful laboratory for testing new physics. The development of bound-state QED has enabled predictions of fundamental spectroscopic quantities with relative precisions as high as $3 \times 10^{-15}$ for an optical transition~\cite{Mohr2025} and $2.8 \times 10^{-11}$ for the bound-electron  $g$ factor \cite{PhysRevA.95.060501}.
Given the precision that can be achieved in such efforts,  even small corrections  of new physics may yield observable signals. Thus, an exhaustive study of the possible interactions and signatures of NP models is warranted to chart the current and future reach within the NP parameter space.

In particular, in highly charged ions (HCI), the strong attractive Coulomb force of the nucleus decreases the size of the electron shell, i.e., the distances between the bound electrons and the nucleons as well as amongst the electrons are smaller. This enhances the sensitivity to shorter range forces and therefore mediators with larger mass ~\cite{Kozlov2018,Debierre2020,Debierre2021,Rehbehn2021,Debierre2022,Akulov2025}. 
This enhancement was already exploited in quantum logic spectroscopy experiments with highly charged Ca ions~\cite{Wilzewski2025}, in studies of the hyperfine structure in HCIs \cite{Quint2025}, and with $g$-factor isotope shift measurements~\cite{Sailer2022,Moretti2026}. In the current article, 
we provide one of the first constraints on electron–electron scalar interactions using bound-state QED observables.
We exploit that the nuclear Coulomb field in such systems also enhances the {\it mutual} interaction of electrons via a new boson, and put forward using precision spectroscopy combined with a fit to multiple measurements as a tool to probe different combinations of electron and nucleon couplings in a more global picture. In particular, the systems we consider enable us to constrain new-scalar mediated interelectronic interactions, similar to what is used in~\cite{Cong:2026kuv} in the context of a solution to the $^3\text{He}$ and $^4\text{He}$ ionization energy anomaly \cite{PhysRevLett.134.223001, PhysRevA.111.012817, PhysRevA.103.042809}, and in~\cite{abdullin2026axionexchangecontributionenergylithiumlike} for the case of axion-like particles.

One of the most precisely measured properties of few-electron highly charged ions is the magnetic moment, or the dimensionless $g$ factor characterizing the latter. Its value is of the order of unity, yielding Larmor frequencies, i.e. transition frequencies between Zeeman sub-levels, typically in the 100-GHz microwave range. This holds regardless of the atomic number $Z$, the degree of ionization, angular momentum, or other parameters of the electron shell. The $g$ factors of highly charged ions can nowadays be measured with $10^{-11}$--$10^{-12}$ relative accuracy by means of the continuous Stern-Gerlach effect in Penning traps~\cite{Sturm2011,Morgner2023,Morgner2025}. Therefore, in the current article, we study the effect of a new force-carrying boson exchanged between two electrons as well as between electrons and nucleons on the $g$ factor.
We work in units of $\hbar = c = 1$.

\section{Theory}
We consider a scalar boson $\phi$ with mass $m_{\phi}$ as the mediator of a hypothetical fifth force between electrons.
Using the slash notation $\slashed{a} := a_\mu\gamma^\mu$, the QED Lagrangian with this added interaction reads (cf., e.g.~\cite{Antypas:2022asj}) 
\begin{align}
    \label{eq:full Lagrangian}
    \mathcal{L} &= \overline{\psi}_e(i\slashed{\partial} -e\slashed{A} - m_e)\psi_e -\frac{1}{4}F_{\mu\nu}F^{\mu\nu} \nonumber \\
    &-\frac{1}{2} \partial_\mu \phi \partial^\mu \phi - \frac{1}{2} m_\phi^2\phi^2 \nonumber \\
    &+ y_e\phi\overline{\psi}_e\psi_e + y_p\phi\overline{\psi}_p\psi_p + y_n\phi\overline{\psi}_n\psi_n \, .
\end{align}
Here $\psi_{e,p,n}$ are electron (e), proton (p), neutron (n) fields, $A_\mu$ is the electromagnetic field, and $F_{\mu\nu}$ is the corresponding field strength tensor. In addition to the usual QED vertex (second term in the brackets), we include a new interaction between the new scalar, electrons, and nucleons. We only describe the dynamics of the electrons in the system, as we assume the neutrons and protons to be stationary due to their large mass compared to the electron. Thus, we treat them as auxiliary fields that couple to the electron as external sources.\\
Starting from the above Lagrangian, we can employ the two-time Green's function formalism (TTGF) outlined in Ref.~\cite{Shabaev2002}, to derive the Feynman rules for electrons in an external Coulomb field. Electron lines are denoted as
\begin{align}
    \begin{gathered}
    \begin{fmffile}{propagator-electron}
        \begin{fmfgraph*}(70,50)
            \fmfstraight
            \fmfleft{i1}
            \fmfright{o1}
            \fmf{dbl_plain}{i1,o1}
        \end{fmfgraph*}
    \end{fmffile}
    \end{gathered}
    \quad
    &= \quad \frac{i}{2\pi}S(\omega,\pmb{x},\pmb{y}) \quad \nonumber \\
    &= \quad \frac{i}{2\pi} \sum_n \frac{\psi_n(\pmb{x})\bar{\psi}_n(\pmb{y})}{\omega-\varepsilon_n (1-i0)},
\end{align}
where $\sum_n$ denotes a sum over the whole discrete bound state spectrum of the Dirac-Coulomb equation, as well as an integration over the continuum part. For internal electron lines, we additionally perform an integration $\int_{\omega}$. The scalar is not bound and we can simply use the free propagator
\begin{align}
    \begin{gathered}
    \begin{fmffile}{propagator-scalar}
        \begin{fmfgraph*}(70,50)
            \fmfstraight
            \fmfleft{i1}
            \fmfright{o1}
            \fmf{dashes}{i1,o1}
        \end{fmfgraph*}
    \end{fmffile}
    \end{gathered}
    \label{eq:scalar propagator}
    \quad
    &= \quad \Delta(\omega,\pmb{x}-\pmb{y}) \quad \nonumber \\
    &= \quad \int \frac{\text{d}^3k}{(2\pi)^3}\, \frac{e^{i\pmb{k}\cdot(\pmb{x}-\pmb{y})}}{\omega^2 - \pmb{k}^2 - m_{\phi}^2 +i0}.
\end{align}
In analogy to the usual Feynman rules, we also obtain an electron-scalar vertex. As the spatial translation invariance is broken by the presence of the nucleus, only energy is conserved at the vertex, i.e.,
\begin{equation}
    \begin{gathered}
    \begin{fmffile}{vertex}
        \begin{fmfgraph*}(70,30)
            \fmfstraight
            \fmfleft{i1,i2}
            \fmfright{o1,o2}
             \fmf{phantom_arrow}{i1,v1}
            \fmf{phantom_arrow}{o1,v1} 
            \fmf{phantom}{i2,v2,o2}
            \fmf{dbl_plain}{i1,v1,o1}  
            \fmffreeze
            \fmf{dashes_arrow}{v2,v1}
            
            \fmfv{decor.shape=circle,decor.size=.15cm,l=$z$}{v1}   
        \end{fmfgraph*}
    \end{fmffile}
    \end{gathered}
    \quad
    = \quad -2\pi i \delta(\omega_1 + \omega_2 + \omega_3) \int \text{d}^3 \pmb{z}.
\end{equation}

Finally, we denote the insertion of an external potential sourced by the nucleus via the exchange of the new scalar boson in question.
\begin{equation}
    \begin{gathered}
    \begin{fmffile}{potential}
        \begin{fmfgraph*}(70,30)
            \fmfstraight
            \fmfleft{i1,i2}
            \fmfright{o1,o2}
            \fmf{phantom_arrow}{i1,v1}
            \fmf{phantom_arrow}{v1,o1} 
            \fmf{phantom}{i2,v2,o2}
            \fmf{dbl_plain}{i1,v1,o1}  
            \fmffreeze
            \fmf{dashes}{v2,v1}
            \fmfv{decor.shape=square,decor.size=.15cm}{v2} 
            \fmfv{decor.shape=circle,decor.size=.15cm,l=$z$}{v1}   
        \end{fmfgraph*}
    \end{fmffile}
    \end{gathered}
    \quad
    = \quad -2\pi \gamma^0 i \delta(\omega_1 -\omega_2) \int \text{d}^3 \pmb{z}\, V(\pmb{z}).\\[1em]
    \nonumber
\end{equation}

\begin{figure}
        \begin{subfigure}{0.49\columnwidth}
            \begin{fmffile}{lilike-noloop-two}
                \begin{fmfgraph*}(60,40)
                    \fmfstraight
                    \fmfleft{i1,i2,i3} 
                    \fmfright{o1,o2,o3}
                    \fmffreeze
                    \fmf{phantom}{i1,v11,v12,o1}
                    \fmf{dbl_plain}{i2,v21,v22,o2} 
                    \fmf{dbl_plain}{i3,v31,v32,o3}
                    \fmffreeze
                    \fmf{photon}{v21,v11}
                    \fmfv{decor.shape=triangle,decor.size=.2cm,decor}{v11}   
                    \fmfv{decor.shape=circle,decor.size=.12cm}{v21}
                    \fmfv{decor.shape=circle,decor.size=.12cm}{v22}
                    \fmf{dashes}{v22,v12}
                    \fmflabel{1s}{i2}
                    \fmflabel{2s}{i3}
                    \fmfv{decor.shape=square,decor.size=.15cm}{v12}
                \end{fmfgraph*}
            \end{fmffile}
            \caption{}   
            \label{fig:Nucleus-electron g-factor 1s Li-like}
        \end{subfigure}
    \begin{subfigure}{0.49\columnwidth}
        \begin{fmffile}{lilike-noloop-one}
            \begin{fmfgraph*}(60,40)
            \fmfstraight
            \fmfleft{i1,i2,i3} 
            \fmfright{o1,o2,o3}
            \fmffreeze
            \fmf{phantom}{i1,v11,v12,o1}
            \fmf{dbl_plain}{i2,v21,v22,o2} 
            \fmf{dbl_plain}{i3,v31,v32,o3}
            \fmffreeze
            \fmf{photon}{v21,v11}
            \fmfv{decor.shape=triangle,decor.size=.2cm,decor}{v11}   
            \fmf{dashes}{v22,v12}
            \fmflabel{1s}{i3}
            \fmflabel{2s}{i2}
            \fmfv{decor.shape=square,decor.size=.15cm}{v12}
            \fmfv{decor.shape=circle,decor.size=.12cm}{v21}
            \fmfv{decor.shape=circle,decor.size=.12cm}{v22}
            \end{fmfgraph*}
        \end{fmffile}
        \caption{}   
        \label{fig:Nucleus-electron g-factor 2s Li-like}
        \end{subfigure}
    \begin{subfigure}{0.49\columnwidth}
        \begin{fmffile}{lilike-loop-2}
            \begin{fmfgraph*}(60,40)
            \fmfstraight
            \fmfleft{i1,i2,i3} 
            \fmfright{o1,o2,o3}
            \fmffreeze
            \fmf{phantom}{i1,v11,v12,v13,v14,o1}
            \fmf{dbl_plain}{i2,v21,v22,v23,v24,o2} 
            \fmf{dbl_plain}{i3,v31,v32,v33,v34,o3}
            \fmffreeze
            \fmf{photon}{v21,v11}
            \fmfv{decor.shape=triangle,decor.size=.2cm,decor}{v11}   
            \fmf{dashes,right=1,tension=.3,dash_len=1}{v22,v24}
            \fmfv{decor.shape=circle,decor.size=.12cm}{v21}
            \fmfv{decor.shape=circle,decor.size=.12cm}{v22}
            \fmfv{decor.shape=circle,decor.size=.12cm}{v24}
            \fmflabel{1s}{i2}
            \fmflabel{2s}{i3}
            \end{fmfgraph*}
        \end{fmffile}
        \caption{}   
        \label{fig:g-factor SE 1s Li-like}
    \end{subfigure}
        \begin{subfigure}{0.49\columnwidth}
            \begin{fmffile}{lilike-loop-1}
                \begin{fmfgraph*}(60,40)
                    \fmfstraight
                    \fmfleft{i1,i2,i3}
                    \fmfright{o1,o2,o3}
                    \fmffreeze
                    \fmf{phantom}{i1,v11,v12,v13,o1}
                    \fmf{dbl_plain}{i2,v21}
                    \fmf{dbl_plain}{v23,o2}
                    \fmf{dbl_plain}{i3,o3}
                    \fmf{plain}{v21,v22,v23}
                    \fmffreeze
                    \fmf{photon}{v22,v12}
                    \fmf{dashes,left}{v21,v23}
                    \fmflabel{2s}{i2}
                    \fmflabel{1s}{i3}
                    \fmfv{decor.shape=triangle,decor.size=.2cm,decor}{v12}   
                    \fmfv{decor.shape=circle,decor.size=.12cm}{v21}
                    \fmfv{decor.shape=circle,decor.size=.12cm}{v22}
                    \fmfv{decor.shape=circle,decor.size=.12cm}{v23}
                \end{fmfgraph*}
            \end{fmffile}
            \caption{}   
            \label{fig:g-factor SE 2s Li-like}
        \end{subfigure}
    \begin{subfigure}{0.49\columnwidth}
        \begin{fmffile}{lilike-iei}
            \begin{fmfgraph*}(60,40)
            \fmfstraight
            \fmfleft{i1,i2,i3} 
            \fmfright{o1,o2,o3}
            \fmffreeze
            \fmf{phantom}{i1,v11,v12,o1}
            \fmf{dbl_plain}{i2,v21,v22,o2} 
            \fmf{dbl_plain}{i3,v31,v32,o3}
            \fmffreeze
            \fmf{photon}{v21,v11}
            \fmfv{decor.shape=triangle,decor.size=.2cm,decor}{v11} 
            \fmfv{decor.shape=circle,decor.size=.12cm,decor}{v21}
            \fmfv{decor.shape=circle,decor.size=.12cm,decor}{v22}
            \fmfv{decor.shape=circle,decor.size=.12cm,decor}{v32}
            \fmf{dashes}{v22,v32}
            \fmflabel{1s}{i3}
            \fmflabel{2s}{i2}
            \end{fmfgraph*}
        \end{fmffile}
        \caption{}
        \label{fig:g-factor one Yukawa}
    \end{subfigure}
    \caption{\justifying Exemplary Feynman diagrams depicting contributions to the bound-electron $g$-factor to leading order in the new physics interaction.
    A double line represents an electron propagating in the Coulomb potential sourced by the nucleus. The state of the electron is indicated on the left of the double line. A wavy line terminated by a triangle denotes an interaction with the external magnetic field (that is, an insertion of the corresponding Zeeman potential), the dashed line terminated by a square an insertion of the new scalar potential sourced by the nucleus, while the dashed line by itself represents the scalar propagator.}
    \label{fig:g-factor all together}
\end{figure}
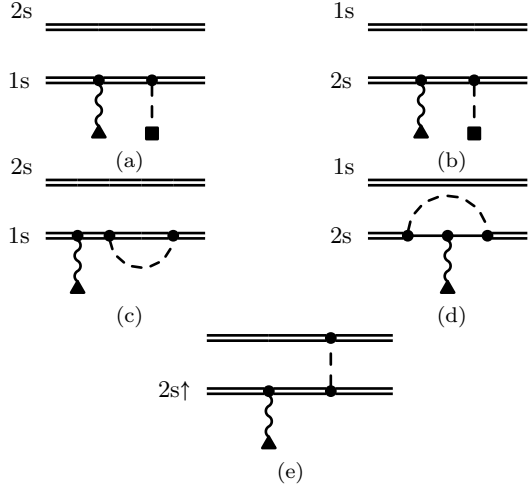

These insertions of a new scalar potential or propagator can modify the $g$ factor of a lithium-like ion via different interactions and corresponding processes: nucleus-electron interactions (Figs.~\ref{fig:Nucleus-electron g-factor 1s Li-like},\ref{fig:Nucleus-electron g-factor 2s Li-like}), electron self energy (SE) contributions (Figs.~\ref{fig:g-factor SE 1s Li-like},\ref{fig:g-factor SE 2s Li-like}) and inter-electronic exchange (Fig.~\ref{fig:g-factor one Yukawa}). In a lithium-like ion, the two electrons in the state $1s$ have opposite magnetic quantum numbers, i.e., $s=\pm \frac{1}{2}$, as a consequence, an external magnetic field coupling to them yields the same contribution but with opposite sign, which is why the two contributions sum up to zero, for the considered model. Hence diagrams of Figs.~\ref{fig:Nucleus-electron g-factor 1s Li-like},\ref{fig:g-factor SE 1s Li-like} do not contribute to the $g$ factor, leaving us with three relevant terms.\footnote{We have also checked that an additional ``disconnected'' diagram where the magnetic interaction and the new physics interaction with the nucleus connect to two different electrons vanishes.} The total $g$ factor shift $\Delta g_\phi$ is then \\
\begin{equation}
    \label{eq:g-factor shift total}
    \Delta g_\phi = \Delta g_\phi^N + \Delta g_\phi^{\text{SE}} + \Delta g_\phi^{\text{inter}},
\end{equation}
where $\Delta g_\phi^N$ refers to Fig.~\ref{fig:Nucleus-electron g-factor 2s Li-like}, $\Delta g_\phi^{\text{SE}}$ to Fig.~\ref{fig:g-factor SE 2s Li-like}, and $\Delta g_\phi^{\text{inter}}$ to Fig~\ref{fig:g-factor one Yukawa}. In the following section, we present the expressions for each one of these three contributions.

\subsection{Electron-nucleus interaction}
\label{subsec:electron-nuclear interaction}
The $2s$ electron interacts with the nucleus via the exchange of the new scalar boson. Taking the nucleus to be nonrelativistic, we obtain a Yukawa potential of the form
\begin{equation}
    \label{eq:Yukawa potential}
    V_\phi(r) =  \alpha_{eN} \tilde{N} \frac{e^{-m_\phi|\bm{r}|}}{|\bm{r}|},
\end{equation}
where the subscript $e$ refers to the electron and $N \in \{n,p\}$ to the nucleon (neutron or proton), and $\tilde{N}$ to the number of those respective nucleons. We further define the reduced coupling constant $\alpha_{eN} = \frac{y_e y_N}{4\pi}$.

A calculation shortcut to obtain $\Delta g_\phi^{eN}$ is to take the derivative of the expectation value of the potential \eqref{eq:Yukawa potential} with respect to the electron mass~\cite{Karshenboim2005}, i.e., the derivative of the energy shift due to the scalar exchange with the nucleus
\begin{equation}
    \label{eq:g-factor electron-nucleus raw}
    \Delta g_\phi^{eN} = \frac{4}{3} \frac{\partial}{\partial m_e} \langle V_\phi\rangle_{2s}.
\end{equation}
For a $2s$ electron we find the following expression for the energy shift induced by the new scalar with a pointlike nucleus

\begin{equation}
    \label{eq:energy vev RQM}
    \langle V_\phi \rangle_{2s} = \pm \, A \left[ B \, \Gamma (2\gamma) + C \, \Gamma (2\gamma +1) + D \, \Gamma (2\gamma + 2) \right] \, ,
\end{equation}
with the following definition of the constants
\begin{align}
    \notag
    A &= \alpha_{eN} N \frac{\lambda}{\Gamma(2\gamma + 1)} \frac{2\gamma + 1}{\eta (\eta + 1)} h(m_{\phi})^{-2\gamma} \, , \notag \\[0.5em]
    \notag
    B &= \eta + 2 \, , \\[0.5em]
    \notag
    C &= -\frac{(\eta + 1)(\eta + 2)}{2\gamma+1} h(m_{\phi})^{-1} \, , \\[0.5em]
    \notag
    D &= \left( \frac{\eta + 1}{2\gamma + 1} \right)^2 h(m_{\phi})^{-2} \, , \nonumber \\[0.5em]
    \notag
    \eta &= \sqrt{2(\gamma + 1)}  \, ,
\end{align}
where we further defined $h(m_{\phi}) = 1+\frac{m_{\phi}}{2\lambda}$.
This way, the mass-derivative of Eq.~\eqref{eq:g-factor electron-nucleus raw} yields
\begin{align}
    \notag
    \Delta g_\phi^{eN} & = \frac{4}{3} \frac{A}{m_e} \Bigg\{ \left[ 1 + \gamma \frac{m_\phi}{\lambda}h(m_{\phi})^{-1} \right] B \, \Gamma (2\gamma) \\
    \notag
    & + \left[ 1 + \left( \gamma + \frac{1}{2} \right) \frac{m_\phi}{\lambda}h(m_{\phi})^{-1} \right] C \, \Gamma (2\gamma +1) \\
    \label{eq:g-factor electron-nucleus RQM}
    & + \left[ 1 + (\gamma + 1) \frac{m_\phi}{\lambda}h(m_{\phi})^{-1} \right] D \, \Gamma (2\gamma + 2)  \Bigg\} \, .
\end{align}
Eqs.~\eqref{eq:energy vev RQM} and \eqref{eq:g-factor electron-nucleus RQM} have been obtained within a relativistic quantum mechanics framework. Equivalent formulae can be derived by employing the TTGF formalism presented above in a quantum field theoretical approach. With this method, the energy shift appears exactly as Eq.~\eqref{eq:energy vev RQM}.  The $g$ factor shift for an electron in the state $a$ is given by
\begin{align}
    \notag
    & \Delta g_\phi^{eN}= -2 \, \alpha_{eN} \, N \, m_a \frac{\kappa_a}{j_a(j_a+1)}\sum_n^{\varepsilon_n\neq\varepsilon_a} \frac{1}{\varepsilon_a - \varepsilon_n} \\
    \notag
    & \quad \times \int_0^\infty \hspace{-1mm} dr \,  \left[ G_{n_a,\kappa_a}F_{n_n,\kappa_a} + F_{n_a,\kappa_a} G_{n_n,\kappa_a} \right](r) \, r \\
    \label{eq:g-factor electron nucleus TTGF}
    & \quad \times \int_0^\infty \hspace{-1mm} dr \, \left[G_{n_a,\kappa_a} G_{n_n,\kappa_a} + F_{n_a,\kappa_a} F_{n_n,\kappa_a} \right] (r)\, \frac{e^{-m_\phi r}}{r} \, ,
\end{align}
where the sum is extended to every possible state $n$ non degenerate with $a$.
This latter equation is equivalent to Eq.~\eqref{eq:g-factor electron-nucleus RQM} and we have checked that it produces the same numeric result, as one would expect. However, it makes it possible to directly include corrections from the finite nuclear size.

\subsection{Self-energy contribution}
\label{subsec:SE interaction}
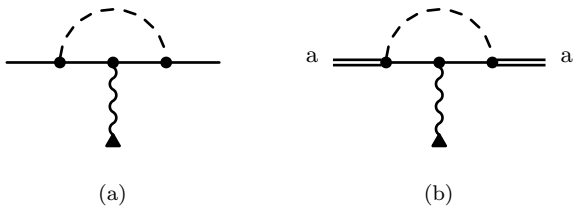
\begin{figure}[b]
\vspace{.8cm}
    \begin{subfigure}{0.33\linewidth}
        \begin{fmffile}{SE-tree}
            \begin{fmfgraph*}(80,30)
            \fmfstraight
                \fmfleft{i1,i2}
                \fmfright{o1,o2}
                \fmffreeze
                \fmf{phantom}{i1,v11,v12,v13,o1}
                \fmf{plain}{i2,v21,v22,v23,o2}
                \fmffreeze
                \fmf{photon}{v22,v12}
                \fmf{dashes,left}{v21,v23}
                \fmfv{decor.shape=triangle,decor.size=.2cm,decor}{v12}   
                \fmfv{decor.shape=circle,decor.size=.12cm}{v21}
                \fmfv{decor.shape=circle,decor.size=.12cm}{v22}
                \fmfv{decor.shape=circle,decor.size=.12cm}{v23}
            \end{fmfgraph*}
        \end{fmffile}\\[1em]
        \caption{}
        \label{fig:g-factor SE free}
    \end{subfigure}
    \hspace{4em}
    \begin{subfigure}{0.33\linewidth}
        \begin{fmffile}{SE-NLO}
            \begin{fmfgraph*}(80,30)
            \fmfstraight
                \fmfleft{i1,i2}
                \fmfright{o1,o2}
                \fmffreeze
                \fmf{phantom}{i1,v11,v12,v13,o1}
                \fmf{dbl_plain}{i2,v21}
                \fmf{dbl_plain}{v23,o2}
                \fmf{plain}{v21,v22,v23}
                \fmffreeze
                \fmf{photon}{v22,v12}
                \fmf{dashes,left}{v21,v23}
                \fmflabel{a}{i2}
                \fmflabel{a}{o2}
                \fmfv{decor.shape=triangle,decor.size=.2cm,decor}{v12}   
                \fmfv{decor.shape=circle,decor.size=.12cm}{v21}
                \fmfv{decor.shape=circle,decor.size=.12cm}{v22}
                \fmfv{decor.shape=circle,decor.size=.12cm}{v23}
            \end{fmfgraph*}
        \end{fmffile}\\[1em]
        \caption{}
        \label{fig:g-factor SE 2s first approx}
    \end{subfigure}
    \caption{\justifying Zeroth- (left) and leading-order (right) terms in the $Z\alpha$ expansion of the one-loop correction to the QED vertex for a bound electron in the state $a$.}
\end{figure}

The computation of the SE correction to the bound-electron g factor is rather involved, and for our purposes, an approximate treatment of this effect will suffice. Our strategy is to treat the SE up to order $(Z\alpha)^2$, since at this order, the heaviest ion we consider has $Z=50$, amounting to a $\sim 1\permil$ correction compared to the free-electron value (see Eq.~\eqref{eq:Zeeman SE 2s first approx}). To do so, we treat the internal electron propagator in Fig.~\ref{fig:g-factor SE 2s Li-like} as a free electron propagator, thus reducing the computation to the evaluation of the diagram in Fig.~\ref{fig:g-factor SE 2s first approx}. As explained in, e.g., Refs.~\cite{Hegstrom1973,Cakir2020}, this corresponds to computing the expectation value of the electron anomalous magnetic moment Hamiltonian $H = a_e^\phi \mu_B \gamma^0 \bm{B\cdot\Sigma}$ with respect to a $2s$ state
\begin{equation}
\label{eq:Zeeman SE 2s first approx}
    \bra{\psi_{2s}}H\ket{\psi_{2s}} = \mu_B B_z \, a_{e}^\phi \left[ 1-\frac{\left(Z\alpha\right)^2}{24} + O\left(\left(Z\alpha\right)^4\right) \right] \, , 
\end{equation}
with $\mu_B$ being the Bohr magneton, $B_z$ the projection of the magnetic potential onto the $z$-axis, and $a_e^\phi$ the anomalous magnetic moment induced by the scalar. The anomalous magnetic moment $a_e^\phi$ is obtained from the evaluation of the diagram in Fig.~\ref{fig:g-factor SE free}, which is the one-loop correction to the QED vertex $-ie\gamma^\mu$ induced by the new scalar field. For any general correction to the QED vertex, it can be given  in the form
\begin{equation}
    \label{eq:vertex function}
    -i e \, \Gamma^\mu(p',p) = -i e \left[ F_1(q^2) \gamma^\mu + \frac{i}{2m}F_2(q^2) \sigma^{\mu\nu} q_\nu \right] \, ,
\end{equation}
with $\sigma^{\mu\nu} = \frac{i}{2}\left[ \gamma^\mu, \gamma^\nu \right]$, when sandwiched between two free Dirac states of momenta $p'$ and $p$, and contracted with a photon of polarization $\epsilon^\mu(q)$. $F_1(q^2)$ and $F_2(q^2)$ are electron form factors accounting for the changes induced by the different types of corrections. While $F_1$ is a renormalization of the electric charge, $F_2$ leads to a modification of the magnetic moment, thus the free electron $g$ factor becomes $g^{\text{free}} = 2 \left( 1+F_2(0)\right) = 2 + \Delta g^{\text{free}}$. Therefore we have $a_{e}=F_{2}(0)$. The contribution to the bound electron $g$ factor, up to orders of $\left( Z\alpha \right)^2$, from the one-loop scalar exchange is then
\begin{equation}
    \label{eq:g-factor SE corrected}
    \Delta g_\phi^{\text{SE}} = 2F_2^\phi(0) \left( 1 - \frac{\left( Z\alpha \right)^2}{24} \right) \, .
\end{equation}
The second form factor, at vanishing transferred momentum $q^2=0$ reads~\cite{Moretti2026}
\begin{widetext}
\begin{equation}
    F_2^\phi (0) = \frac{\alpha_{ee}}{2\pi} \left\{ \frac{3}{2} - r_m^2 -\frac{r_m \left( r_m^4 - 5r_m^2 +4 \right)}{\sqrt{r_m^2 - 4}} \left[ \text{atanh} \left( \frac{r_m^2 - 2}{r_m\sqrt{r_m^2 - 4}} \right) - \text{atanh} \left( \frac{r_m}{\sqrt{r_m^2 - 4}} \right) \right] + r_m^2 \left( r_m^2 - 3 \right) \text{ln}\left( r_m \right) \right\} \, ,
\end{equation}
\end{widetext}
where $r_m=m_\phi/m_e$ is the ratio of scalar and electron mass, and $\alpha_{ee} = y_e^2/4\pi$.

\subsection{Inter-electronic interaction}
\label{subsec:inter-electronic interaction}
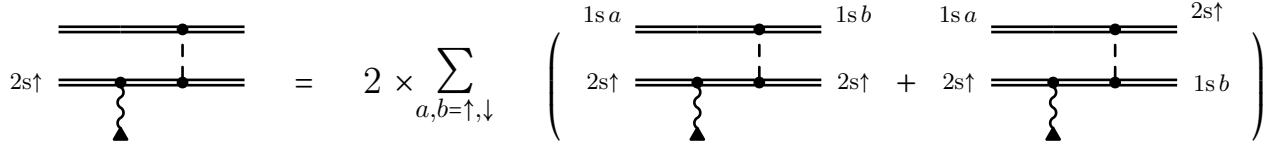
\begin{figure*}
\begin{equation*}
    \begin{gathered}
       \begin{fmffile}{lilike-iei}
            \begin{fmfgraph*}(70,40)
            \fmfstraight
            \fmfleft{i1,i2,i3} 
            \fmfright{o1,o2,o3}
            \fmffreeze
            \fmf{phantom}{i1,v11,v12,o1}
            \fmf{dbl_plain}{i2,v21,v22,o2} 
            \fmf{dbl_plain}{i3,v31,v32,o3}
            \fmffreeze
            \fmf{photon}{v21,v11}
            \fmfv{decor.shape=triangle,decor.size=.2cm,decor}{v11}
            \fmfv{decor.shape=circle,decor.size=.12cm,decor}{v21}
            \fmfv{decor.shape=circle,decor.size=.12cm,decor}{v22}
            \fmfv{decor.shape=circle,decor.size=.12cm,decor}{v32}
            \fmf{dashes}{v22,v32}
            \fmflabel{}{i3}
            \fmflabel{2s$\uparrow$}{i2}
        \end{fmfgraph*}
        \end{fmffile}
    \end{gathered}
    \quad\quad \mathlarger{\mathlarger{\mathlarger{=}}} \quad\quad
    \mathlarger{\mathlarger{\mathlarger{2 \,\,\times}}}
    \begin{gathered}
 \mathlarger{\mathlarger{\mathlarger{\sum_{a,b=\uparrow,\downarrow}}}}
   \end{gathered}
   \quad\quad
   \left( \quad \quad \quad
   \begin{gathered}
       \begin{fmffile}{lilike-iei-explicit-1}
            \begin{fmfgraph*}(70,40)
            \fmfstraight
            \fmfleft{i1,i2,i3} 
            \fmfright{o1,o2,o3}
            \fmffreeze
            \fmf{phantom}{i1,v11,v12,o1}
            \fmf{dbl_plain}{i2,v21,v22,o2} 
            \fmf{dbl_plain}{i3,v31,v32,o3}
            \fmffreeze
            \fmf{photon}{v21,v11}
            \fmfv{decor.shape=triangle,decor.size=.2cm,decor}{v11}  
            \fmfv{decor.shape=circle,decor.size=.12cm,decor}{v21}
            \fmfv{decor.shape=circle,decor.size=.12cm,decor}{v22}
            \fmfv{decor.shape=circle,decor.size=.12cm,decor}{v32} 
            \fmf{dashes}{v22,v32}
            \fmflabel{1s$\,a$}{i3}
            \fmflabel{2s$\uparrow$}{i2}
            \fmflabel{1s$\,b$}{o3}
            \fmflabel{2s$\uparrow$}{o2}
        \end{fmfgraph*}
    \end{fmffile}   
    \end{gathered}
    \quad\quad\quad \mathlarger{\mathlarger{\mathlarger{+}}}
    \quad\quad\quad
    \begin{gathered}
       \begin{fmffile}{lilike-iei-explicit-2}
            \begin{fmfgraph*}(70,40)
            \fmfstraight
            \fmfleft{i1,i2,i3} 
            \fmfright{o1,o2,o3}
            \fmffreeze
            \fmf{phantom}{i1,v11,v12,o1}
            \fmf{dbl_plain}{i2,v21,v22,o2} 
            \fmf{dbl_plain}{i3,v31,v32,o3}
            \fmffreeze
            \fmf{photon}{v21,v11}
            \fmfv{decor.shape=triangle,decor.size=.2cm,decor}{v11}   
            \fmfv{decor.shape=circle,decor.size=.12cm,decor}{v21}
            \fmfv{decor.shape=circle,decor.size=.12cm,decor}{v22}
            \fmfv{decor.shape=circle,decor.size=.12cm,decor}{v32}
            \fmf{dashes}{v22,v32}
            \fmflabel{1s$\,a$}{i3}
            \fmflabel{2s$\uparrow$}{i2}
            \fmflabel{1s$\,b$}{o2}
            \fmflabel{2s$\uparrow$}{o3}
        \end{fmfgraph*}
    \end{fmffile}   
    \end{gathered}\quad\quad\quad
    \right)
\end{equation*}
    \caption{\justifying Sum over all exchange contributions, with two additional factors $2$ accounting for the possible change of the order of interactions and for the choice of the electron interacting with the external magnetic field.}
    \label{fig:g-factor-one-scalar-all}
\end{figure*}

The last correction to the $g$ factor of a lithium-like ion, at first order in the scalar interaction, stems from the exchange of one scalar particle between two electrons, as illustrated in the example diagram~\ref{fig:g-factor one Yukawa}.
In total there are 16 diagrams of this type (see Fig.~\ref{fig:g-factor-one-scalar-all}): we have a factor of 2 depending on which electron interacts with the external magnetic field, another factor of 2 for the exchange of the electrons in the final state, a further factor of 2 for the exchange of the order of magnetic and inter-electronic interactions, and a final factor of 2 because the inner shell contains two $1s$ electrons, with spin equal and opposite to the shell electron spin. We can see that a subset of the diagrams in Fig.~\ref{fig:g-factor-one-scalar-all} gives a vanishing contribution because the scalar, being spinless, does not change the spin of the particles it interacts with. Also, exchanging the order of magnetic and scalar interactions produces the very same result.

To compute these diagrams, we employ the TTGF formalism~\cite{Shabaev2002}, which separates each diagram into an irreducible and a reducible part, and summed in the general form $\Delta g = \Delta g^{\text{irr}}+\Delta g^{\text{red}}$. For each diagram we write, 
\begin{equation}
    \label{eq:g-factor inter-electronic red+irr}
    \Delta g_\phi^{\text{inter}} = \Delta g_\phi^{\text{inter,red}} + \Delta g_\phi^{\text{inter,irr}} \, .
\end{equation}
The two contributions can be expressed as
\begin{align}
    \notag
    & \Delta g_\phi^{\text{inter,irr}} = 2 \sum_c \Big( \bra{\delta v,c}\Delta(0)\ket{v,c} + \bra{v,\delta c}\Delta(0)\ket{v,c} \\
    \label{eq:g-factor inter-electronic irr}
    & \quad\quad - \bra{\delta v,c}\Delta(\varepsilon_c - \varepsilon_v)\ket{c,v} - \bra{v,\delta c}\Delta(\varepsilon_c - \varepsilon_v)\ket{c,v} \Big) \, , \\
    \notag
    & \Delta g_\phi^{\text{inter,red}} = \sum_c \bra{v,c}\Delta'(\varepsilon_c - \varepsilon_v)\ket{c,v} \\
    \label{eq:g-factor inter-electronic red}
    & \quad\quad\quad\quad \quad\quad\quad\quad\times \Big( \bra{c}V_{\text{mag}}\ket{c} - \bra{v}V_{mag}\ket{v} \Big) \, ,
\end{align}
where $v$ and $c$, respectively, stand for valence and core states, $\Delta$ is the scalar propagator introduced in Eq.~\eqref{eq:scalar propagator}  as function of the energy level difference of the pre- and post-interaction unperturbed Dirac states, $\Delta'(\varepsilon)$ denotes its derivative with respect to $\varepsilon$, and a perturbed state $\ket{\delta a}$ is the first-order perturbative correction to the bound-electron wave function
\begin{equation}
    \ket{\delta a} = \sum_n^{\varepsilon_n \neq \varepsilon_a} \ket{n} \frac{\bra{n}V_{\text{mag}}\ket{a}}{\varepsilon_a - \varepsilon_n} \, .
\end{equation}
From this, the inter-electronic interaction contribution becomes readily calculable after angular reduction, i.e., carrying out the integrals over spherical harmonics involved in this calculation. The calculation of the matrix element of $\Delta$ is presented in App.~\ref{app:one-scalar exchange}.

\begin{figure}
    \centering
    \includegraphics[width=\linewidth]{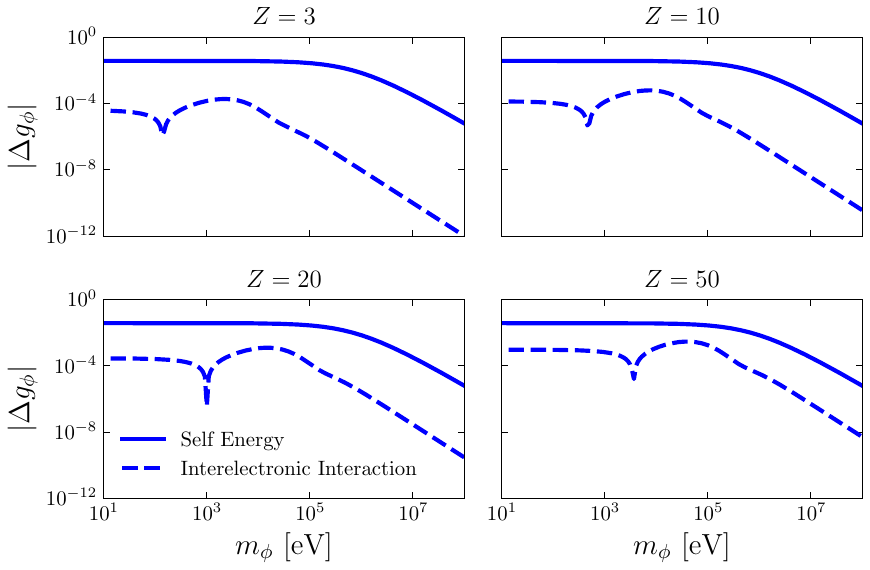}
    \caption{\justifying Comparison of self-energy (solid lines) and inter-electronic (dashed lines) corrections to the $g$ factor for a few ions, for the value of the electron-scalar coupling $y_e = 1$. As the figure shows, for increasing values of the atomic number $Z$ the inter-electronic contribution grows larger becoming more and more significant.}
    \label{fig:SE vs inter}
\end{figure}

Although, for light ions, the inter-electronic contribution is very small compared to the self-energy contribution, this is no longer the case for heavy ions. This is because the distance between electrons decreases as $Z$ increases. Fig.~\ref{fig:SE vs inter} illustrates the increasing impact of the inter-electronic contribution on the $g$ factor with increasing atomic number $Z$. In lithium-like ions, the two contributions therefore need to be considered, as they become closer in size for larger values of $Z$.

\section{Results \& Data Analysis}
In addition to the Li-like $g$-factor observables, we can use other, very precise, measurements to constrain the couplings in question. In particular, we introduce hydrogen-like neon, isotope shift and free-electron $g$-factor constraints as additional data points. 

In principle, new scalar bosons coupling to electrons and protons may also affect the determined value of the fine-structure constant $\alpha$ -- and thus to the free-electron $g$ factor~\cite{Volkov2019}.  The reason is that the value of $\alpha$, determined via photon recoil, depends on the Rydberg constant obtained from transitions in the hydrogen atom~\cite{CODATA2014}, where a new electron-proton coupling might be present. However, the error budget of $\alpha$ is dominated by systematic and statistical errors and the uncertainty of $R_\infty$ plays a minor role~\cite{Parker2018,Bouchendira2011}. Therefore, we can neglect this source of possible correlation between $y_e y_p$ and $y_e y_e$.

\begin{figure*}
    \centering
    \includegraphics[width=\linewidth]{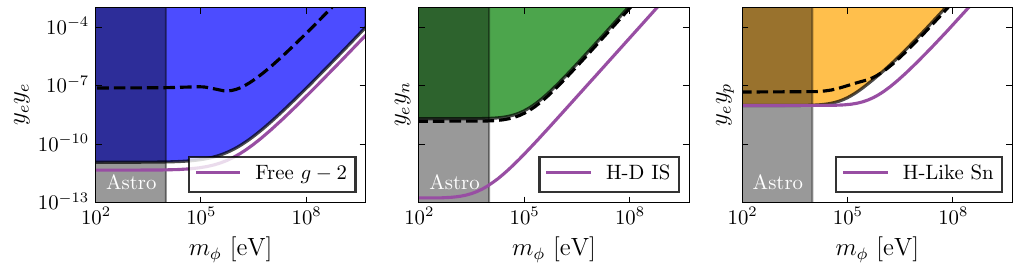}
    \caption{\justifying $95 \%$ C.I. constraints on the couplings from the $\chi^2$ fit are shown as filled areas enclosed by black, solid lines. Solid purple lines indicate standalone constraints from Refs.\ \cite{Fan2023,Delaunay:2017dku,Moretti2026} which we show for comparison . Limits without using the free electron $g-2$ are shown as dashed lines.}
    \label{fig:contr_single}
\end{figure*}

\begin{figure*}
    \centering
    \includegraphics[width=\linewidth]{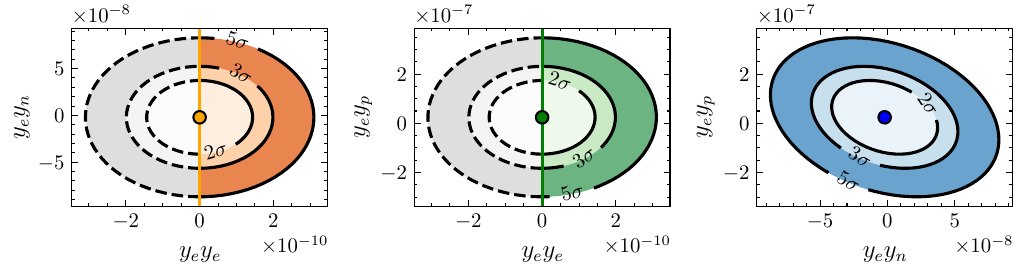}
    \caption{\justifying Exclusion contours at $m_{\phi}/m_e = 1$ for all three combinations of couplings. We give the combined results in orange, green, and blue, and the data that constitutes them in the indicated shades. For the $y_ey_n/y_ey_p$ parameter space, we see a slight anti-correlation.}
    \label{fig:corr}
\end{figure*}

\begin{figure*}
    \centering
    \includegraphics[width=\linewidth]{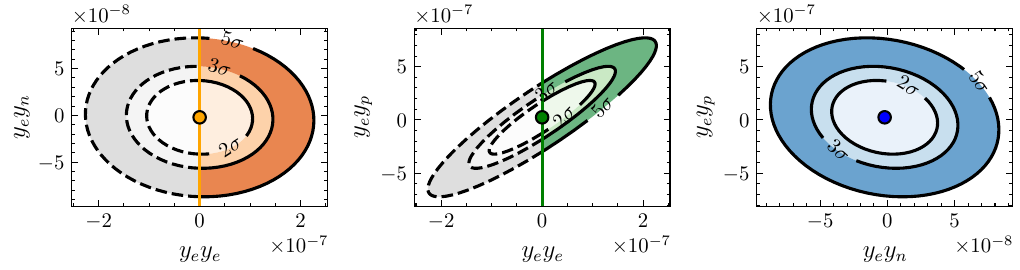}
    \caption{\justifying Exclusion contours at $m_{\phi}/m_e = 1$ for all three combinations of couplings without the free electron $g$-factor. We give the combined results in orange, green, and blue, and the data that constitutes them in the indicated shades. For the $y_ey_n/y_ey_p$ parameter space, we see a slight anti-correlation, and a slight correlation for the $y_ey_e/yey_p$ parameter space.}
    \label{fig:corr no free g-2}
\end{figure*}

The results are given in two ways, we first show the different coupling vs.\ mass parameter spaces in Fig.~\ref{fig:contr_single}.  
Second, in Figs.~\ref{fig:corr},\ \ref{fig:corr no free g-2} we show two dimensional constraints for different coupling combinations, for an exemplary mass value. 

In the plots, it becomes clear that the most dominant constraint is still from the free electron $g-2$.

The contributions due to different coupling combinations entering the $g$ factor given in Eq.~\eqref{eq:g-factor shift total} are indistinguishable within the same measurement. We therefore combine measurements on different systems. To do so we proceed similarly to the method proposed in Ref.~\cite{Jaeckel2010} and perform a $\chi^2$-fit to multiple measurements. In order for the system of equations not to be undetermined, we consider multiple observables $m_i$, which depend on the three couplings $y_e y_j = 4\pi \, \alpha_{ej}$ and $m_\phi$ with $j \in \{e,n,p\}$. In the following, we use the observables listed in \autoref{tab:obs}. 

\begin{table}[h]
\begin{ruledtabular}
\begin{tabular}{l|c|c|c}
     {\rm Observable} & $x_i=|g_{\rm{Th}}-g_{\rm{Exp}}|$ & $\sigma_i$ & Refs.\   \\
    \hline
     $g(\text{Ne}^{9+})$ & $1.20 \times 10^{-10}$ & $2.25 \times 10^{-10}$ & \cite{Heisse2023} \\[.3em]
     $g(\text{Si}^{11+})$ & $5.1 \times 10^{-9}$ & $1.9 \times 10^{-8}$ & \cite{Wagner2013} \\[.3em]
     $g(\text{Ca}^{17+})$ & $1.5 \times 10^{-9}$ & $1.30 \times 10^{-8}$ & \cite{Koehler2016} \\[.3em]
     $g(\text{Sn}^{47+})$ & $2.8 \times 10^{-9}$ & $1.20 \times 10^{-8}$ & \cite{Morgner2025} \\[.3em]
     $g(^{20}\text{Ne}^{9+})-g(^{22}\text{Ne}^{9+})$& $1.2 \times 10^{-12}$ & $1.11 \times 10^{-11}$ & \cite{Sailer2022} \\[.3em]
     $g(e^-)$& $1.38 \times 10^{-12}$ & $7.3 \times 10^{-13}$ & \cite{Volkov2019,Fan2023} \\[.3em]
\end{tabular}
\end{ruledtabular}
\caption{\justifying The observables we choose, the corresponding differences between SM theory and experiment, along with the associated uncertainty. The references are given in the last column.}
\label{tab:obs}
\end{table}

We then obtain our unconstrained $\chi^2$ in matrix form (cf.\ the statistics section of Ref.~\cite{ParticleDataGroup:2024cfk}) as

\begin{equation}
    \chi^2(y_{e}y_j) = (\pmb{x}- \pmb{\mu}(y_{e}y_j))^T V^{-1}(\pmb{x}-\pmb{\mu}(y_{e}y_j)), 
\end{equation}
where $\pmb{x}$ denotes the vector of differences between theory and experiment for the different observables given in \autoref{tab:obs} (without accounting for uncertainties), $\pmb{\mu}(y_{e}y_j)$ is the vector of contributions of the new physics coupling to the different observables with each entry corresponding to an observable $m_i$, and $V$ is the covariance, which is given by
\begin{equation}
    V_{ij} = \rho_{ij}\sigma_i\sigma_j,
\end{equation}
with the errors given by the experimental and theoretical errors as $\sigma_i = \sqrt{(\sigma_i^{\text{exp}})^2+(\sigma_i^{\text{theo}})^2}$. 
For simplicity we simply assume the different measurements to be uncorrelated, i.e. $\rho_{ij} = \delta_{ij}$. Considering we are investigating mostly bound-electron $g$-factors, this is obviously an oversimplification and a next step would be to carefully determine correlations arising from the input data of the theory calculation and the experimental extraction of the $g$ factor. Despite this we can still get insights into potential degeneracies in the directions of the new physics couplings.

At the leading order that we have calculated the new physics contributions have the simple form
\begin{equation}
    \mu_k(y_{e}y_j) = y_ey_j H_{kj},
\end{equation}
where $H$ is a $N\times3$-dimensional matrix, with $N$ being the number of observables, containing the numerical values of the linearized new physics contributions as a function of the new boson mass. Its columns correspond to different couplings, and the rows to the different observables.  From this, the best fit values for the new physics couplings are obtained via 
\begin{equation}
    y_ey_j = (H^T V^{-1}H)^{-1}_{jl}H^T_{lm}V^{-1}_{mn}x_n\,.
\end{equation}
From this, we can readily fit our three couplings to existing experimental and theoretical data.

We obtain a covariance in the space of couplings, given by 
\begin{equation}
    C^{-1} = H^T V^{-1} H.
\end{equation}
from this, we can obtain correlations in the space of couplings, that is, 
\begin{equation}
    \text{corr}(y_ey_j,y_ey_i) = \frac{C_{ij}}{\sqrt{C_{ii}\cdot C_{ij}}}.
\end{equation}
In principle this has a non-trivial dependence on the correlation matrix $V^{-1}$ of the used data. Here, we nevertheless stick with our simplifying assumption of this being uncorrelated and focus on the new physics contribution. 

In general the $g$ factor measurements with different ions are somewhat similar and therefore leave a partial degeneracy. However, the electron $g$ factor and the isotope shift between the different neon ions focus on the electron and neutron coupling, respectively, helping to lift the degeneracy.
We further implement a positivity constraint onto the quadratic coupling $y_ey_e$ . This optimization problem is solved by finding a set of couplings that solves the Karush-Kuhn-Tucker (KKT) conditions (cf.\ Ref.\ \cite{Kuhn1951}). Practically speaking, we first perform an unconstrained fit, and if the resulting best fit value for $y_ey_e$ is negative, we enforce an equality constraint $y_ey_e = 0$ by formulating the optimization with the Lagrangian
\begin{equation}
    \mathcal{L}(y_ey_j, \lambda) = (\pmb{x}- \pmb{\mu}(y_{e}y_j))^T V^{-1}(\pmb{x}-\pmb{\mu}(y_{e}y_j)) + \pmb{\lambda}^T \pmb{\theta}\,.
\end{equation}
The new matrix equation then becomes
\begin{equation}
    \begin{pmatrix}
        F & L^T \\
        L & \pmb{0}
    \end{pmatrix}
    \begin{pmatrix}
        \pmb{\theta} \\
        \pmb{\lambda}
    \end{pmatrix}
    = 
    \begin{pmatrix}
        \pmb{S} \\
        0
    \end{pmatrix},
\end{equation}
where $F = H^T V^{-1} H$, $S = H^T V^{-1}\pmb{x}$, and $L= \pmb{e}_1$.

 The constraints on the separate parameter spaces are shown in Fig.\ \ref{fig:contr_single}, and the ellipses in the coupling-coupling spaces are shown in Fig.\ \ref{fig:corr}.
We observe the expected feature that spectroscopy bounds supersede astrophysical constraints on new physics for masses $\gtrsim 10 \, \text{keV}$, and decay as a power law for higher masses due to the Yukawa suppression of the interaction.

Fig.\ \ref{fig:corr no free g-2} shows our results with the free electron $g$ factor excluded from the data. This demonstrates that the considered couplings can be constrained with $g$ factor measurements alone. However, noting the different scale on the axis we can also see that this leads to a significantly reduced sensitivity, since all coupling combinations contain the electron coupling which is highly constrained by the electron $g$ factor measurement.

\section{Concluding remarks}
The $g$ factors of few-electron  highly charged ions are nowadays routinely measured with amazing precision, with theory keeping up in making similarly precise predictions. These measurements therefore offer themselves as a natural testbed to probe new physics models.\footnote{As an observation, we note that the $g$ factor contributions due to the electron coupling arise from topologically distinct diagrams which, for more complicated types of effective interactions may have a different behavior and corresponding weighting between them. This could potentially increase the discriminating power when adding $g$ factor measurements.} In this work we have taken first steps to realize this potential and used existing data to constrain a new scalar coupled to electrons, protons and neutrons. Notably, as data exists for different ion species we can go beyond simplistic ``one coupling combination'' assumptions and constrain all possible products of the three couplings simultaneously. 
While highly charged ion $g$ factors can constrain all coupling combinations, the limiting power is significantly enhanced by including the free-electron $g$ factor, which is highly sensitive to the
electron coupling.
Importantly, the small spatial extent of the highly charged ions enables probing masses into the MeV-GeV range and enters regions unconstrained by current astrophysical limits.

At present the g-factor measurements we use provide bounds roughly comparable to the best single coupling combination limits obtained by other methods. Already a relatively mild reduction in the QED theoretical uncertainties by a factor of 3 for Li-like systems would improve the sensitivity to be competitive with the current best limit for $y_ey_p$. Such an improvement of bound-state QED theory is plausible in the foreseeable future. Under the additional assumption that the relevant experimental and theoretical uncertainties could be reduced further, g-factor spectroscopy could in principle provide constraints beyond current bounds, while retaining the increased model-independence provided by including multiple coupling combinations and using spectroscopic data. Such improvements would require substantial but realistic progress beyond the present level of theoretical and experimental precision.

\emph{Acknowledgments} — Supported by the Deutsche Forschungsgemeinschaft (DFG, German Research Foundation) Project-ID 273811115—SFB 1225 IsoQuant.

\bibliographystyle{apsrev4-2}
\bibliography{bibliography}

@misc{Jaeckel:2026aeh,
    author = "Jaeckel, Joerg and Puetter, Lucas",
    title = "{Testing Antimatter Couplings with Spectroscopy}",
    eprint = "2608.07656",
    archivePrefix = "arXiv",
    primaryClass = "hep-ph",
    month = "8",
    year = "2026"
}

@article{Antypas:2022asj,
    author = "Antypas, D. and others",
    title = "{New Horizons: Scalar and Vector Ultralight Dark Matter}",
    eprint = "2203.14915",
    archivePrefix = "arXiv",
    primaryClass = "hep-ex",
    reportNumber = "FERMILAB-PUB-22-262-AD-PPD-T",
    month = "3",
    journal="",
    year = "2022"
}

@misc{Delaunay:2026ymq,
      title={Atomic Spectroscopy Probes of New Physics}, 
      author={Cédric Delaunay and Jean-Philippe Karr and Yotam Soreq},
      year={2026},
      eprint={2602.20750},
      archivePrefix={arXiv},
      primaryClass={hep-ph},
      url={https://arxiv.org/abs/2602.20750}, 
}

@article{Potvliege:2023lvf,
    author = "Potvliege, Robert M. and Nicolson, Adair and Jones, Matthew P. A. and Spannowsky, Michael",
    title = "{Deuterium spectroscopy for enhanced bounds on physics beyond the standard model}",
    eprint = "2309.03732",
    archivePrefix = "arXiv",
    primaryClass = "hep-ph",
    reportNumber = "IPPP/23/50",
    doi = "10.1103/PhysRevA.108.052825",
    journal = "Phys. Rev. A",
    volume = "108",
    number = "5",
    pages = "052825",
    year = "2023"
}

@article{Shabaev2002,
  title = "Two-time Green's function method in quantum electrodynamics of high-Z few-electron atoms",
  journal = "Phys. Rep.",
  volume = "356",
  number = "3",
  pages = "119 - 228",
  year = "2002",
  issn = "0370-1573",
  doi = "https://doi.org/10.1016/S0370-1573(01)00024-2",
  url = "http://www.sciencedirect.com/science/article/pii/S0370157301000242",
  author = "V. M. Shabaev"
}

@article{Karshenboim2005,
  title = {$g$ factor of an electron or muon bound by an arbitrary central potential},
  author = {Karshenboim, S. G. and Lee, R. N. and Milstein, A. I.},
  journal = {Phys. Rev. A},
  volume = {72},
  issue = {4},
  pages = {042101},
  numpages = {5},
  year = {2005},
  month = {Oct},
  publisher = {American Physical Society},
  doi = {10.1103/PhysRevA.72.042101},
  url = {https://link.aps.org/doi/10.1103/PhysRevA.72.042101}
}

@article{Hegstrom1973,
  title = {Nuclear-Mass and Anomalous-Moment Corrections to the Hamiltonian for an Atom in a Constant External Magnetic Field},
  author = {Hegstrom, Roger A.},
  journal = {Phys. Rev. A},
  volume = {7},
  issue = {2},
  pages = {451--456},
  numpages = {0},
  year = {1973},
  month = {Feb},
  publisher = {American Physical Society},
  doi = {10.1103/PhysRevA.7.451},
  url = {https://link.aps.org/doi/10.1103/PhysRevA.7.451}
}

@article{Sturm2011,
  title = {$g$ Factor of Hydrogenlike $^{28}\mathrm{Si}^{13+}$},
  author = {Sturm, S. and Wagner, A. and Schabinger, B. and Zatorski, J. and Harman, Z. and Quint, W. and Werth, G. and Keitel, C. H. and Blaum, K.},
  journal = {Phys. Rev. Lett.},
  volume = {107},
  issue = {2},
  pages = {023002},
  numpages = {4},
  year = {2011},
  month = {Jul},
  publisher = {American Physical Society},
  doi = {10.1103/PhysRevLett.107.023002},
  url = {https://link.aps.org/doi/10.1103/PhysRevLett.107.023002}
}

@article{Koehler2016,
  title = {Isotope dependence of the Zeeman effect in lithium-like calcium},
  author = {K\"{o}hler, Florian and Blaum, Klaus and Block, Michael and Chenmarev, Stanislav and Eliseev, Sergey and
  Glazov, Dmitry A. and Goncharov, Mikhail and Hou, Jiamin and Kracke, Anke and Nesterenko, Dmitri A. and Novikov, Yuri N. and
  Quint, Wolfgang and Minaya Ramirez, Enrique and Shabaev, Vladimir M. and Sturm, Sven and Volotka, Andrey V. and Werth, G{\"u}nter},
  journal = {Nat. Commun.},
  volume = {7},
  issue = {1},
  pages = {10246},
  year = {2016},
  doi = {10.1038/ncomms10246},
  url = {https://doi.org/10.1038/ncomms10246}
}

@article{Wagner2013,
  title = {$g$ Factor of Lithiumlike Silicon $^{28}\mathrm{Si}^{11\mathbf{+}}$},
  author = {Wagner, A. and Sturm, S. and K\"ohler, F. and Glazov, D. A. and Volotka, A. V. and Plunien, G. and Quint, W. and Werth, G. and Shabaev, V. M. and Blaum, K.},
  journal = {Phys. Rev. Lett.},
  volume = {110},
  issue = {3},
  pages = {033003},
  numpages = {5},
  year = {2013},
  month = {Jan},
  publisher = {American Physical Society},
  doi = {10.1103/PhysRevLett.110.033003},
  url = {https://link.aps.org/doi/10.1103/PhysRevLett.110.033003}
}

@article{Heisse2023,
  title = {{High-Precision Determination of $g$ Factors and Masses of ${^{20}\mathrm{Ne}}^{9+}$ and ${^{22}\mathrm{Ne}}^{9+}$}},
  author = {Hei\ss{}e, F. and Door, M. and Sailer, T. and Filianin, P. and Herkenhoff, J. and K\"onig, C. M. and Kromer, K. and Lange, D. and Morgner, J. and Rischka, A. and Schweiger, Ch. and Tu, B. and Novikov, Y. N. and Eliseev, S. and Sturm, S. and Blaum, K.},
  journal = {Phys. Rev. Lett.},
  volume = {131},
  issue = {25},
  pages = {253002},
  numpages = {7},
  year = {2023},
  month = {Dec},
  publisher = {American Physical Society},
  doi = {10.1103/PhysRevLett.131.253002},
  url = {https://link.aps.org/doi/10.1103/PhysRevLett.131.253002}
}

@article{Sailer2022,
  title = {Measurement of the bound-electron $g$-factor difference in coupled ions},
  author = {Sailer, T. and Debierre, V. and Harman, Z. and Heiße, F. and König, C. and Morgner, J. and Tu, B. and Volotka, A. V. and Keitel, C. H. and Blaum, K. and Sturm, S.},
  journal = {Nature},
  volume = {606},
  pages = {479-483},
  year = {2022},
  month = {Jun},
  issue = {7914},
  publisher = {Nature},
  doi = {10.1038/s41586-022-04807-w},
  url = {https://doi.org/10.1038/s41586-022-04807-w}
}

@article{Morgner2023,
title = {Stringent test of {QED} with hydrogen-like tin},
  author = {Morgner, J. and Tu, B. and Koenig C. M. and Sailer, T. and Heisse, F. and Bekker, H and Sikora, B. and Lyu, C. and Yerokhin, V. A. and Harman, Z. and {Crespo López-Urrutia}, J. R. and Keitel, C. H. and Sturm, S. and Blaum, K.},
  journal = {Nature},
  volume = {622},
  pages = {53-57},
  year = {2023},
  month = {Oct},
  issue = {7981},
  publisher = {Nature},
  doi = {10.1038/s41586-023-06453-2},
  url = {https://doi.org/10.1038/s41586-023-06453-2}
}

@article{Cakir2020,
  title = {{QED} corrections to the $g$ factor of {L}i- and {B}-like ions},
  author = {Cakir, H. and Yerokhin, V. A. and Oreshkina, N. S. and Sikora, B. and Tupitsyn, I. I. and Keitel, C. H. and Harman, Z.},
  journal = {Phys. Rev. A},
  volume = {101},
  issue = {6},
  pages = {062513},
  numpages = {14},
  year = {2020},
  month = {Jun},
  publisher = {American Physical Society},
  doi = {10.1103/PhysRevA.101.062513},
  url = {https://link.aps.org/doi/10.1103/PhysRevA.101.062513}
}

@article{Volkov2019,
  title = {Calculating the five-loop {QED} contribution to the electron anomalous magnetic moment: {G}raphs without lepton loops},
  author = {Volkov, S.},
  journal = {Phys. Rev. D},
  volume = {100},
  issue = {9},
  pages = {096004},
  numpages = {13},
  year = {2019},
  month = {Nov},
  publisher = {American Physical Society},
  doi = {10.1103/PhysRevD.100.096004},
  url = {https://link.aps.org/doi/10.1103/PhysRevD.100.096004}
}

@article{Fan2023,
  title = {Measurement of the Electron Magnetic Moment},
  author = {Fan, X. and Myers, T. G. and Sukra, B. A. D. and Gabrielse, G.},
  journal = {Phys. Rev. Lett.},
  volume = {130},
  issue = {7},
  pages = {071801},
  numpages = {6},
  year = {2023},
  month = {Feb},
  publisher = {American Physical Society},
  doi = {10.1103/PhysRevLett.130.071801},
  url = {https://link.aps.org/doi/10.1103/PhysRevLett.130.071801}
}

@article{Debierre2020,
author = {Debierre, Vincent and Keitel, C. H. and Harman, Zoltán},
year = {2020},
month = {06},
pages = {135527},
title = {Fifth-force search with the bound-electron $g$ factor},
volume = {807},
journal = {Phys. Lett. B},
doi = {10.1016/j.physletb.2020.135527}
}

@article{Door2025,
  title = {Probing New Bosons and Nuclear Structure with Ytterbium Isotope Shifts},
  author = {Door, Menno and Yeh, Chih-Han and Heinz, Matthias and Kirk, Fiona and Lyu, Chunhai and Miyagi, Takayuki and Berengut, Julian C. and Biero\ifmmode \acute{n}\else \'{n}\fi{}, Jacek and Blaum, Klaus and Dreissen, Laura S. and Eliseev, Sergey and Filianin, Pavel and Filzinger, Melina and Fuchs, Elina and F\"urst, Henning A. and Gaigalas, Gediminas and Harman, Zolt\'an and Herkenhoff, Jost and Huntemann, Nils and Keitel, Christoph H. and Kromer, Kathrin and Lange, Daniel and Rischka, Alexander and Schweiger, Christoph and Schwenk, Achim and Shimizu, Noritaka and Mehlst\"aubler, Tanja E.},
  journal = {Phys. Rev. Lett.},
  volume = {134},
  issue = {6},
  pages = {063002},
  numpages = {7},
  year = {2025},
  month = {Feb},
  publisher = {American Physical Society},
  doi = {10.1103/PhysRevLett.134.063002},
  url = {https://link.aps.org/doi/10.1103/PhysRevLett.134.063002}
}

@article{Jones2020,
  title = {Probing new physics using Rydberg states of atomic hydrogen},
  author = {Jones, Matthew P. A. and Potvliege, Robert M. and Spannowsky, Michael},
  journal = {Phys. Rev. Res.},
  volume = {2},
  issue = {1},
  pages = {013244},
  numpages = {16},
  year = {2020},
  month = {Mar},
  publisher = {American Physical Society},
  doi = {10.1103/PhysRevResearch.2.013244},
  url = {https://link.aps.org/doi/10.1103/PhysRevResearch.2.013244}
}

@article{Potvliege2025,
  title={Spectroscopy of light atoms and bounds on physics beyond the standard model},
  author={Potvliege, R. M.},
  journal={New J. Phys.},
  volume={27},
  number={4},
  pages={045002},
  year={2025},
  publisher={IOP Publishing}
}

@article{Moretti2026,
  title = {Fermion-Selective Tests of New Physics with the Bound-Electron $g$ Factor},
  author = {Moretti, M. and Keitel, C. H. and Harman, Z.},
  journal = {Phys. Rev. Lett.},
  volume = {136},
  issue = {1},
  pages = {011803},
  numpages = {6},
  year = {2026},
  month = {Jan},
  publisher = {American Physical Society},
  doi = {10.1103/zyb6-lvy8},
  url = {https://link.aps.org/doi/10.1103/zyb6-lvy8}
}

@article{Wilzewski2025,
  title = {Nonlinear Calcium King Plot Constrains New Bosons and Nuclear Properties},
  author = {Wilzewski, Alexander and Spie\ss{}, Lukas J. and Wehrheim, Malte and Chen, Shuying and King, Steven A. and Micke, Peter and Filzinger, Melina and Steinel, Martin R. and Huntemann, Nils and Benkler, Erik and Schmidt, Piet O. and Huber, Luca I. and Flannery, Jeremy and Matt, Roland and Stadler, Martin and Oswald, Robin and Schmid, Fabian and Kienzler, Daniel and Home, Jonathan and Craik, Diana P. L. Aude and Door, Menno and Eliseev, Sergey and Filianin, Pavel and Herkenhoff, Jost and Kromer, Kathrin and Blaum, Klaus and Yerokhin, Vladimir A. and Valuev, Igor A. and Oreshkina, Natalia S. and Lyu, Chunhai and Banerjee, Sreya and Keitel, Christoph H. and Harman, Zolt\'an and Berengut, Julian C. and Viatkina, Anna and Gilles, Jan and Surzhykov, Andrey and Rosner, Michael K. and Crespo L\'opez-Urrutia, Jos\'e R. and Richter, Jan and Mariotti, Agnese and Fuchs, Elina},
  journal = {Phys. Rev. Lett.},
  volume = {134},
  issue = {23},
  pages = {233002},
  numpages = {10},
  year = {2025},
  month = {Jun},
  publisher = {American Physical Society},
  doi = {10.1103/PhysRevLett.134.233002},
  url = {https://link.aps.org/doi/10.1103/PhysRevLett.134.233002}
}

@article{Debierre2021,
  title = {Radiative and photon-exchange corrections to new-physics contributions to energy levels in few-electron ions},
  author = {Debierre, V. and Oreshkina, N. S.},
  journal = {Phys. Rev. A},
  volume = {104},
  issue = {3},
  pages = {032825},
  numpages = {11},
  year = {2021},
  month = {Sep},
  publisher = {American Physical Society},
  doi = {10.1103/PhysRevA.104.032825},
  url = {https://link.aps.org/doi/10.1103/PhysRevA.104.032825}
}

@article{Debierre2022,
  title = {Testing standard-model extensions with isotope shifts in few-electron ions},
  author = {Debierre, V. and Oreshkina, N. S. and Valuev, I. A. and Harman, Z. and Keitel, C. H.},
  journal = {Phys. Rev. A},
  volume = {106},
  issue = {6},
  pages = {062801},
  numpages = {13},
  year = {2022},
  month = {Dec},
  publisher = {American Physical Society},
  doi = {10.1103/PhysRevA.106.062801},
  url = {https://link.aps.org/doi/10.1103/PhysRevA.106.062801}
}

@article{Quint2025,
  title = {Stringent Constraints on New Pseudoscalar and Vector Bosons from Precision Hyperfine Splitting Measurements},
  author = {Quint, Cedric and Hei\ss{}e, Fabian and Jaeckel, Joerg and Leimenstoll, Lutz and Keitel, Christoph H. and Harman, Zolt\'an},
  journal = {Phys. Rev. Lett.},
  volume = {136},
  issue = {11},
  pages = {113001},
  numpages = {9},
  year = {2026},
  month = {Mar},
  publisher = {American Physical Society},
  doi = {10.1103/rvt1-93v2},
  url = {https://link.aps.org/doi/10.1103/rvt1-93v2}
}

@Article{Akulov2025,
AUTHOR = {Akulov, Dmitry S. and Abdullin, Rinat R. and Chubukov, Dmitry V. and Glazov, Dmitry A. and Volotka, Andrey V.},
TITLE = {{$g$-Factor Isotopic Shifts: Theoretical Limits on New Physics Search}},
JOURNAL = {Atoms},
VOLUME = {13},
YEAR = {2025},
NUMBER = {6},
ARTICLE-NUMBER = {52},
URL = {https://www.mdpi.com/2218-2004/13/6/52},
ISSN = {2218-2004},
DOI = {10.3390/atoms13060052}
}

@article{Counts2020,
  title = {Evidence for Nonlinear Isotope Shift in ${\mathrm{Yb}}^{+}$ Search for New Boson},
  author = {Counts, Ian and Hur, Joonseok and Aude Craik, Diana P. L. and Jeon, Honggi and Leung, Calvin and Berengut, Julian C. and Geddes, Amy and Kawasaki, Akio and Jhe, Wonho and Vuleti\ifmmode \acute{c}\else \'{c}\fi{}, Vladan},
  journal = {Phys. Rev. Lett.},
  volume = {125},
  issue = {12},
  pages = {123002},
  numpages = {7},
  year = {2020},
  month = {Sep},
  publisher = {American Physical Society},
  doi = {10.1103/PhysRevLett.125.123002},
  url = {https://link.aps.org/doi/10.1103/PhysRevLett.125.123002}
}

@article{Rehbehn2021,
  title = {Sensitivity to new physics of isotope-shift studies using the coronal lines of highly charged calcium ions},
  author = {Rehbehn, Nils-Holger and Rosner, Michael K. and Bekker, Hendrik and Berengut, Julian C. and Schmidt, Piet O. and King, Steven A. and Micke, Peter and Gu, Ming Feng and M\"uller, Robert and Surzhykov, Andrey and L\'opez-Urrutia, Jos\'e R. Crespo},
  journal = {Phys. Rev. A},
  volume = {103},
  issue = {4},
  pages = {L040801},
  numpages = {7},
  year = {2021},
  month = {Apr},
  publisher = {American Physical Society},
  doi = {10.1103/PhysRevA.103.L040801},
  url = {https://link.aps.org/doi/10.1103/PhysRevA.103.L040801}
}

@article{Berengut2018,
  title = {Probing New Long-Range Interactions by Isotope Shift Spectroscopy},
  author = {Berengut, Julian C. and Budker, Dmitry and Delaunay, C\'edric and Flambaum, Victor V. and Frugiuele, Claudia and Fuchs, Elina and Grojean, Christophe and Harnik, Roni and Ozeri, Roee and Perez, Gilad and Soreq, Yotam},
  journal = {Phys. Rev. Lett.},
  volume = {120},
  issue = {9},
  pages = {091801},
  numpages = {7},
  year = {2018},
  month = {Feb},
  publisher = {American Physical Society},
  doi = {10.1103/PhysRevLett.120.091801},
  url = {https://link.aps.org/doi/10.1103/PhysRevLett.120.091801}
}

@article{Safronova2018,
  title = {Search for new physics with atoms and molecules},
  author = {Safronova, M. S. and Budker, D. and DeMille, D. and Kimball, Derek F. Jackson and Derevianko, A. and Clark, Charles W.},
  journal = {Rev. Mod. Phys.},
  volume = {90},
  issue = {2},
  pages = {025008},
  numpages = {106},
  year = {2018},
  month = {Jun},
  publisher = {American Physical Society},
  doi = {10.1103/RevModPhys.90.025008},
  url = {https://link.aps.org/doi/10.1103/RevModPhys.90.025008}
}

@article{Kozlov2018,
  title = {Highly charged ions: Optical clocks and applications in fundamental physics},
  author = {Kozlov, M. G. and Safronova, M. S. and Crespo L\'opez-Urrutia, J. R. and Schmidt, P. O.},
  journal = {Rev. Mod. Phys.},
  volume = {90},
  issue = {4},
  pages = {045005},
  numpages = {49},
  year = {2018},
  month = {Dec},
  publisher = {American Physical Society},
  doi = {10.1103/RevModPhys.90.045005},
  url = {https://link.aps.org/doi/10.1103/RevModPhys.90.045005}
}

@article{Flambaum2018,
  title = {Isotope shift, nonlinearity of {King} plots, and the search for new particles},
  author = {Flambaum, V. V. and Geddes, A. J. and Viatkina, A. V.},
  journal = {Phys. Rev. A},
  volume = {97},
  issue = {3},
  pages = {032510},
  numpages = {12},
  year = {2018},
  month = {Mar},
  publisher = {American Physical Society},
  doi = {10.1103/PhysRevA.97.032510},
  url = {https://link.aps.org/doi/10.1103/PhysRevA.97.032510}
}

@article{Delaunay2017,
  title = {Probing atomic Higgs-like forces at the precision frontier},
  author = {Delaunay, C\'edric and Ozeri, Roee and Perez, Gilad and Soreq, Yotam},
  journal = {Phys. Rev. D},
  volume = {96},
  issue = {9},
  pages = {093001},
  numpages = {7},
  year = {2017},
  month = {Nov},
  publisher = {American Physical Society},
  doi = {10.1103/PhysRevD.96.093001},
  url = {https://link.aps.org/doi/10.1103/PhysRevD.96.093001}
}

@article{Jaeckel2010,
  title = {{Spectroscopy as a test of Coulomb's law: A probe of the hidden sector}},
  author = {Jaeckel, Joerg and Roy, Sabyasachi},
  journal = {Phys. Rev. D},
  volume = {82},
  issue = {12},
  pages = {125020},
  numpages = {13},
  year = {2010},
  month = {Dec},
  publisher = {American Physical Society},
  doi = {10.1103/PhysRevD.82.125020},
  url = {https://link.aps.org/doi/10.1103/PhysRevD.82.125020}
}

@article{JaeckelRingwald2010,
   author = "Jaeckel, Joerg and Ringwald, Andreas",
   title = "The Low-Energy Frontier of Particle Physics",
   journal= "Annual Review of Nuclear and Particle Science",
   year = "2010",
   volume = "60",
   number = "Volume 60, 2010",
   pages = "405-437",
   doi = "https://doi.org/10.1146/annurev.nucl.012809.104433",
   url = "https://www.annualreviews.org/content/journals/10.1146/annurev.nucl.012809.104433",
   publisher = "Annual Reviews",
   issn = "1545-4134",
   type = "Journal Article"
  }

@article{Morgner2025,
author = {Jonathan Morgner  and Vladimir A. Yerokhin  and Charlotte M. König  and Fabian Heiße  and Bingsheng Tu  and Tim Sailer  and Bastian Sikora  and Zoltán Harman  and José R. Crespo López-Urrutia  and Christoph H. Keitel  and Sven Sturm  and Klaus Blaum },
title = {Testing interelectronic interaction in lithium-like tin},
journal = {Science},
volume = {388},
number = {6750},
pages = {945-949},
year = {2025},
doi = {10.1126/science.adn5981},
URL = {https://www.science.org/doi/abs/10.1126/science.adn5981},
eprint = {https://www.science.org/doi/pdf/10.1126/science.adn5981}
}

@article{ParticleDataGroup:2024cfk,
    author = "Navas, S. and others",
    collaboration = "Particle Data Group",
    title = "{Review of particle physics}",
    doi = "10.1103/PhysRevD.110.030001",
    journal = "Phys. Rev. D",
    volume = "110",
    number = "3",
    pages = "030001",
    year = "2024"
}

@article{Ginges2004,
title = {Violations of fundamental symmetries in atoms and tests of unification theories of elementary particles},
journal = {Phys. Rep.},
volume = {397},
number = {2},
pages = {63-154},
year = {2004},
issn = {0370-1573},
doi = {https://doi.org/10.1016/j.physrep.2004.03.005},
url = {https://www.sciencedirect.com/science/article/pii/S0370157304001322},
author = {J.S.M. Ginges and V.V. Flambaum}
}

@article{Rehbehn2023,
  title = {Narrow and Ultranarrow Transitions in Highly Charged Xe Ions as Probes of Fifth Forces},
  author = {Rehbehn, Nils-Holger and Rosner, Michael K. and Berengut, Julian C. and Schmidt, Piet O. and Pfeifer, Thomas and Gu, Ming Feng and L\'opez-Urrutia, Jos\'e R. Crespo},
  journal = {Phys. Rev. Lett.},
  volume = {131},
  issue = {16},
  pages = {161803},
  numpages = {7},
  year = {2023},
  month = {Oct},
  publisher = {American Physical Society},
  doi = {10.1103/PhysRevLett.131.161803},
  url = {https://link.aps.org/doi/10.1103/PhysRevLett.131.161803}
}

@article{Parker2018,
author = {Richard H. Parker  and Chenghui Yu  and Weicheng Zhong  and Brian Estey  and Holger Müller },
title = {Measurement of the fine-structure constant as a test of the Standard Model},
journal = {Science},
volume = {360},
number = {6385},
pages = {191-195},
year = {2018},
doi = {10.1126/science.aap7706},
URL = {https://www.science.org/doi/abs/10.1126/science.aap7706},
eprint = {https://www.science.org/doi/pdf/10.1126/science.aap7706}
}

@article{CODATA2014,
  title = {CODATA recommended values of the fundamental physical constants: 2014},
  author = {Mohr, Peter J. and Newell, David B. and Taylor, Barry N.},
  journal = {Rev. Mod. Phys.},
  volume = {88},
  issue = {3},
  pages = {035009},
  numpages = {73},
  year = {2016},
  month = {Sep},
  publisher = {American Physical Society},
  doi = {10.1103/RevModPhys.88.035009},
  url = {https://link.aps.org/doi/10.1103/RevModPhys.88.035009}
}

@article{Bouchendira2011,
  title = {New Determination of the Fine Structure Constant and Test of the Quantum Electrodynamics},
  author = {Bouchendira, Rym and Clad\'e, Pierre and Guellati-Kh\'elifa, Sa\"{\i}da and Nez, Fran\ifmmode \mbox{\c{c}}\else \c{c}\fi{}ois and Biraben, Fran\ifmmode \mbox{\c{c}}\else \c{c}\fi{}ois},
  journal = {Phys. Rev. Lett.},
  volume = {106},
  issue = {8},
  pages = {080801},
  numpages = {4},
  year = {2011},
  month = {Feb},
  publisher = {American Physical Society},
  doi = {10.1103/PhysRevLett.106.080801},
  url = {https://link.aps.org/doi/10.1103/PhysRevLett.106.080801}
}

@incollection{Kuhn1951,
  author    = {Kuhn, Harold W. and Tucker, Albert W.},
  title     = {Nonlinear Programming},
  editor    = {Jerzy Neyman},
  booktitle = {Proceedings of the Second Berkeley Symposium on Mathematical Statistics and Probability},
  publisher = {University of California Press},
  address   = {Berkeley, CA},
  year      = {1951},
  pages     = {481--492},
  doi       = {10.1525/9780520411586-036}
}

@misc{Cong:2026kuv,
      title={Testing Exotic Electron-Electron Interactions with the Helium Ionization-Energy Anomaly}, 
      author={Lei Cong and Filip Ficek and Rinat Abdullin and Mikhail G. Kozlov and Dmitry Budker},
      year={2026},
      eprint={2602.09743},
      archivePrefix={arXiv},
      primaryClass={physics.atom-ph},
      url={https://arxiv.org/abs/2602.09743}, 
}

@article{PhysRevLett.134.223001,
  title = {Ionization Energy of Metastable $^{3}\mathrm{He}$ (2 $^{3}{S}_{1}$) and the Alpha- and Helion-Particle Charge-Radius Difference from Precision Spectroscopy of the $np$ Rydberg Series},
  author = {Clausen, Gloria and Merkt, Fr\'ed\'eric},
  journal = {Phys. Rev. Lett.},
  volume = {134},
  issue = {22},
  pages = {223001},
  numpages = {9},
  year = {2025},
  month = {Jun},
  publisher = {American Physical Society},
  doi = {10.1103/PhysRevLett.134.223001},
  url = {https://link.aps.org/doi/10.1103/PhysRevLett.134.223001}
}

@article{PhysRevA.111.012817,
  title = {Metrology in a two-electron atom: The ionization energy of metastable triplet helium $2^{3}\mathrm{S}_{1}$},
  author = {Clausen, Gloria and Gamlin, Kai and Agner, Josef A. and Schmutz, Hansj\"urg and Merkt, Fr\'ed\'eric},
  journal = {Phys. Rev. A},
  volume = {111},
  issue = {1},
  pages = {012817},
  numpages = {13},
  year = {2025},
  month = {Jan},
  publisher = {American Physical Society},
  doi = {10.1103/PhysRevA.111.012817},
  url = {https://link.aps.org/doi/10.1103/PhysRevA.111.012817}
}

@article{PhysRevA.103.042809,
  title = {Complete ${\ensuremath{\alpha}}^{7}m$ Lamb shift of helium triplet states},
  author = {Patk\'o\ifmmode \check{s}\else \v{s}\fi{}, Vojt\ifmmode \check{e}\else \v{e}\fi{}ch and Yerokhin, Vladimir A. and Pachucki, Krzysztof},
  journal = {Phys. Rev. A},
  volume = {103},
  issue = {4},
  pages = {042809},
  numpages = {10},
  year = {2021},
  month = {Apr},
  publisher = {American Physical Society},
  doi = {10.1103/PhysRevA.103.042809},
  url = {https://link.aps.org/doi/10.1103/PhysRevA.103.042809}
}

@misc{abdullin2026axionexchangecontributionenergylithiumlike,
      title={Axion-Exchange Contribution to the Energy of Lithium-Like Ions}, 
      author={R. R. Abdullin and A. V. Volotka and D. A. Glazov and M. G. Kozlov and A. D. Moshkin and D. V. Chubukov},
      year={2026},
      eprint={2605.12444},
      archivePrefix={arXiv},
      primaryClass={physics.atom-ph},
      url={https://arxiv.org/abs/2605.12444}, 
}

@article{Delaunay:2017dku,
    author = "Delaunay, C{\'e}dric and Frugiuele, Claudia and Fuchs, Elina and Soreq, Yotam",
    title = "{Probing new spin-independent interactions through precision spectroscopy in atoms with few electrons}",
    eprint = "1709.02817",
    archivePrefix = "arXiv",
    primaryClass = "hep-ph",
    doi = "10.1103/PhysRevD.96.115002",
    journal = "Phys. Rev. D",
    volume = "96",
    number = "11",
    pages = "115002",
    year = "2017"
}

@article{PhysRevA.95.060501,
  title = {One-loop electron self-energy for the bound-electron $g$ factor},
  author = {Yerokhin, V. A. and Harman, Z.},
  journal = {Phys. Rev. A},
  volume = {95},
  issue = {6},
  pages = {060501(R)},
  numpages = {4},
  year = {2017},
  month = {Jun},
  publisher = {American Physical Society},
  doi = {10.1103/PhysRevA.95.060501},
  url = {https://link.aps.org/doi/10.1103/PhysRevA.95.060501}
}

@article{Mohr2025,
  title = {CODATA recommended values of the fundamental physical constants: 2022},
  author = {Mohr, Peter J. and Newell, David B. and Taylor, Barry N. and Tiesinga, Eite},
  journal = {Rev. Mod. Phys.},
  volume = {97},
  issue = {2},
  pages = {025002},
  numpages = {62},
  year = {2025},
  month = {Apr},
  publisher = {American Physical Society},
  doi = {10.1103/RevModPhys.97.025002},
  url = {https://link.aps.org/doi/10.1103/RevModPhys.97.025002}
}

\newpage
\onecolumngrid

\appendix

\section{One-scalar exchange}
\label{app:one-scalar exchange}
In the following we show the explicit calculation of the matrix element $\bra{a,b}\Delta(\omega_m)\ket{c,d}$,
\begin{equation}
\label{eq:one-scalar exchange matrix element}
    \bra{a,b}\Delta(\omega)\ket{c,d} = y_e^2 \int d^3x \, d^3y \, \psi^\dagger_a(\bm{x})\gamma^0\psi_c(\bm{x}) \, \psi^\dagger_b(\bm{y})\gamma^0\psi_d(\bm{y}) \, \Delta(\omega,\bm{x}-\bm{y}) \, ,
\end{equation}
where $\Delta(\omega,\bm{x}-\bm{y})$ is the scalar propagator of Eq.~\eqref{eq:scalar propagator}, that, after integrating out the momentum $k$, can be expanded as
\begin{equation}
    \label{eq:Taylor-Laplace scalar propagator}
    \Delta(\omega,\bm{x}-\bm{y}) = -\omega_m \sum_{l=0}^\infty R_l(\omega_m,r_>,r_<)\sum_{m=-l}^l (-1)^m \, Y_l^m(\bm{\hat{x}}) Y_l^{-m}(\bm{\hat{y}}) \, ,
\end{equation}
with $\omega_m = \sqrt{|\omega^2 - m_\phi^2|}$, $\bm{\hat{x}} = \frac{\bm{x}}{|\bm{x}|}$, and $R_l$ enclosing the radial dependence as function of the maximum, $r_>$, and minimum, $r_<$, of $x$ and $y$ 
\begin{equation}
    \label{eq:Rl}
    R_l(\omega_m,r_>,r_<) = \left\{
    \begin{aligned}
        & i\, h_l^{(1)}(\omega_m \, r_>) \, j_l(\omega_m \, r_<)\,, & \text{if} \quad m_\phi < \omega \\
        & \kappa_l(\omega_m \, r_>) \, i_l(\omega_m \, r_<)\,, & \text{if} \quad m_\phi > \omega
    \end{aligned} \quad ,
    \right.
\end{equation}
and $h_l^{(1)}$, $j_l$, $\kappa_l$, $i_l$ being the radial Hankel function of the first kind, spherical Bessel function of the first kind, modified spherical Bessel function of first and second kind, respectively. Notice that in our case the variable $\omega$ is the energy difference between two states (see Eqs.~\eqref{eq:g-factor inter-electronic irr},~\eqref{eq:g-factor inter-electronic red}). When the scalar mass lies below the available energy, the scalar particle can behave effectively like a photon, leading to an additional imaginary contribution. However, this contribution is canceled by the corresponding SE diagram associated with the same state (see App.~\ref{app:Cutkosky}). Physically, this can be interpreted as the light scalar being able to leave the ion, while an equivalent contribution from virtual scalars in the SE and VP corrections compensates for this effect, resulting in an overall cancellation.\\
Depending on our notation convenience, we label bound-electron states either by $(n,\kappa,m)$ and $(n,j,l,m)$. For a generic state the bound-electron wave function is then
\begin{equation}
    \label{eq:electronic wave function}
    \psi(\bm{x}) = \left(
    \begin{aligned}
        g_{n,\kappa}(x) \, \Omega_{j,l,m}(\bm{\hat{x}})\\
        i \, f_{n,\kappa}(x) \, \Omega_{j,l',m}(\bm{\hat{x}})
    \end{aligned}
    \right) \, ,
\end{equation}
with the bi-spinors
\begin{equation}
    \label{eq:bi-spinor}
    \Omega_{j,l,m}(\bm{\hat{x}}) = \sum_{m_s=\pm1/2} \CG{l}{m-m_s}{1/2}{m_s}{j}{m} \, Y_l^{m-m_s} \, \chi_{\frac{1}{2},m_s} \, ,
\end{equation}
and $\chi_{1/2,m_s}$ being two-component orthonormal vectors, $\chi_{1/2,-1/2} = \left( \begin{aligned} 0\\1 \end{aligned} \right)$, $\chi_{1/2,+1/2} = \left( \begin{aligned} 1\\0 \end{aligned} \right)$.\\
The relations among $j$, $l$, $l'$ and $\kappa$ are
\begin{subequations}
    \begin{align}
        & \kappa = \left( j+1/2 \right)(-1)^{(j+l+1/2)} \, , \label{eq:kappa} \\
        & j = |\kappa|-1/2 \, , \label{eq:j} \\
        & l = |\kappa + 1/2|-1/2 \, , \label{eq:l} \\
        & l' = |-\kappa + 1/2|-1/2 = l\pm1\, . \label{eq:l'}
    \end{align}
\end{subequations}
Clebsch-Gordan coefficients and 3-$j$ symbols are related as 
\begin{equation}
    \label{eq:CG-3j relation}
    \CG{j_1}{m_1}{j_2}{m_2}{j_3}{m_3} = (-1)^{j_1+j_2+j_3}\sqrt{2j_3+1} \threej{j_1}{m_1}{j_2}{m_2}{j_3}{-m_3} \, ,
\end{equation}
and 3-$j$ symbols satisfy the following identities:
\begin{subequations}
    \label{eq:3j properties}
    \begin{align}
        \label{eq:3j even}
        & \quad\quad \threej{j_1}{m_1}{j_2}{m_2}{j_3}{m_3} = \threej{j_2}{m_2}{j_3}{m_3}{j_1}{m_1} = \threej{j_3}{m_3}{j_1}{m_1}{j_2}{m_2} \, ; \\
        \label{eq:3j odd}
        & \quad\quad \threej{j_1}{m_1}{j_2}{m_2}{j_3}{m_3} = (-1)^{j_1+j_2+j_3} \threej{j_1}{m_1}{j_3}{m_3}{j_2}{m_2} = (-1)^{j_1+j_2+j_3} \threej{j_3}{m_3}{j_2}{m_2}{j_1}{m_1} \, ;\\
        \label{eq:3j m's}
        & \quad\quad \threej{j_1}{m_1}{j_2}{m_2}{j_3}{m_3} = (-1)^{j_1+j_2+j_3} \threej{j_1}{-m_1}{j_2}{-m_2}{j_3}{-m_3} \, .
    \end{align}
\end{subequations}
Two useful identities are
\begin{align}
    \label{eq:3j sum to 6j}
    & \sum_{m_1',m_2',m_3'} (-1)^{l_1+l_2+l_3+m_1'+m_2'+m_3'} \threej{j_1}{m_1}{l_2}{m_2'}{l_3}{-m_3'} \threej{l_1}{-m_1'}{j_2}{m_2}{l_3}{m_3'} \threej{l_1}{m_1'}{l_2}{-m_2'}{j_3}{m_3} = \threej{j_1}{m_1}{j_2}{m_2}{j_3}{m_3} \sixj{j_1}{j_2}{j_3}{l_1}{l_2}{l_3} \, ;\\
    \label{eq:3j * 6j}
    & \sqrt{(2l_1+1)(2l_2+1)}\threej{l_1}{0}{l_2}{0}{l}{0}\sixj{j_1}{j_2}{l}{l_2}{l_1}{1/2} = -\threej{j_1}{1/2}{j_2}{-1/2}{l}{0} \Pi(l_1+l_2+l) \, ,
\end{align}
where
\begin{equation}
    \label{eq:pi function}
    \Pi(l)=\left\{ \begin{aligned} &1 &\text{if $l$ is even} \\ &0 &\text{otherwise} \end{aligned} \right.
\end{equation}
Eq.~\eqref{eq:one-scalar exchange matrix element} becomes
{\small
\begin{align}
    \notag
    & \bra{a,b}\Delta(\omega_m)\ket{c,d} = - y_e^2 \omega_m \sum_{l=0}^\infty \int d^3x\,d^3y \, R_l(\omega_m, r_>, r_<) \sum_{m=-l}^l (-1)^m \left(\psi^\dagger_{n_a,j_a,m_a} \gamma^0 \psi_{n_c,j_c,m_c}\right)(\bm{x}) \cdot \left(\psi^\dagger_{n_b,j_b,m_b} \gamma^0 \psi_{n_d,j_d,m_d}\right)(\bm{y}) \\
    \notag
    & = - y_e^2 \, \omega_m \sum_{l=0}^\infty \int d^3x\,d^3y \, R_l(\omega_m, r_>, r_<) \sum_{m=-l}^l (-1)^m  \\
    \notag
    & \quad \times  \left[ G_{n_a,j_a}G_{n_c,j_c}(x)\int d\bm{\hat{x}} \left( \Omega^\dagger_{j_a,l_a,m_a}\Omega_{j_c,l_c,m_c} Y_l^m \right)(\bm{\hat{x}}) - F_{n_a,j_a}F_{n_c,j_c}(x)\int d\bm{\hat{x}} \left( \Omega^\dagger_{j_a,l_a',m_a}\Omega_{j_c,l_c',m_c} Y_l^m \right) (\bm{\hat{x}}) \right] \\ 
    \label{eq:one-scalar exchange stretched}
    & \quad \times \left[ G_{n_b,j_b}G_{n_d,j_d}(y)\int d\bm{\hat{y}} \left( \Omega^\dagger_{j_b,l_b,m_b} \Omega_{j_d,l_d,m_d} Y_l^{-m} \right)(\bm{\hat{y}}) -  F_{n_b,j_b}F_{n_d,j_d}(y)\int d\bm{\hat{y}} \left( \Omega^\dagger_{j_b,l_b',m_b}\Omega_{j_d,l_d',m_d} Y_l^{-m} \right)(\bm{\hat{y}}) \right] \, .
\end{align}}%
Now we focus on the angular integrations, and in particular on the first one of the four of Eq.\eqref{eq:one-scalar exchange matrix element}, since all of them are very similar. Because spherical harmonics are spherical tensor operators, we can use the Wigner-Eckart theorem, for which,
\begin{subequations}
    \label{eq:WEs}
    \begin{align}
        \label{eq:WE general}
        & \bra{j_a,l_a,m_a}Y_l^m\ket{j_c,l_c,m_c} \coloneq \int d\bm{\hat{x}} \left( \Omega^\dagger_{j_a,l_a,m_a} \Omega_{j_c,l_c,m_c} Y_l^m \right) (\bm{\hat{x}}) = (-1)^{j_a-m_a} \threej{j_a}{-m_a}{l}{m}{j_c}{m_c} \langle j_a,l_a\|Y_l\|j_c,l_c\rangle\,, \quad \text{and}\\
        \label{eq:WE specific}
        & \bra{j_a,l_a,1/2}Y_l^0\ket{j_c,l_c,1/2} \coloneq \int d\bm{\hat{x}} \left( \Omega^\dagger_{j_a,l_a,1/2} \Omega_{j_c,l_c,1/2} Y_l^0 \right) (\bm{\hat{x}}) = (-1)^{j_a-1/2} \threej{j_a}{-1/2}{l}{0}{j_c}{1/2} \langle j_a,l_a\|Y_l\|j_c,l_c\rangle \, ,
    \end{align}
\end{subequations}
where $\langle j_a,l_a\|Y_l\|j_c,l_c\rangle$ is the reduced matrix element. Because of the separation of the magnetic quantum numbers, we can restrict ourself in the computation of Eq.~\eqref{eq:WE specific} to find the reduced matrix element, and plug it into Eq.~\eqref{eq:WE general},
\begin{subequations}
    \label{eq:angular integral specific}
    \begin{align}
        \label{eq:angular integral a}
        & \bra{j_a,l_a,1/2}Y_l^0\ket{j_c,l_c,1/2} = \int d\bm{\hat{x}} \left( \Omega^\dagger_{j_a,l_a,1/2} \Omega_{j_c,l_c,1/2} Y_l^0 \right)(\bm{\hat{x}}) \\
        \notag
        & = \int d\bm{\hat{x}} \sum_{m_{s_a},m_{s_c}} \CG{l_a}{1/2-m_{s_a}}{1/2}{m_{s_a}}{j_a}{1/2} \CG{l_c}{1/2-m_{s_c}}{1/2}{m_{s_c}}{j_c}{m_c} \\
        \label{eq:angular integral specific b}
        & \quad\times \left( \left( Y_{l_a}^{1/2-m_{s_a}} \right)^* Y_{l_c}^{1/2-m_{s_c}} Y_l^0 \right) (\bm{\hat{x}}) \, \chi_{1/2,m_{s_a}}^\dagger \chi_{1/2,m_{s_c}} \\
        \notag
        & = \sum_{m_{s_a},m_{s_c}} (-1)^{-1/2+m_{s_a}} \, \delta_{m_{s_a},m_{s_c}} \CG{l_a}{1/2-m_{s_a}}{1/2}{m_{s_a}}{j_a}{1/2} \CG{l_c}{1/2-m_{s_c}}{1/2}{m_{s_c}}{j_c}{1/2} \\
        \label{eq:angular integral specific c}
        & \quad\times \int d\bm{\hat{x}} \left( Y_{l_a}^{-1/2+m_{s_a}} Y_{l_c}^{1/2-m_{s_c}} Y_l^0 \right) (\bm{\hat{x}}) \\ 
        \label{eq:angular integral specific d}
        & = \CG{l_a}{0}{l_c}{0}{l}{0} \sqrt{\frac{(2l_a+1)(2l_c+1)}{4\pi (2l+1)}} \sum_{m_s} \sum_{m_{l_a},m_{l_c}} \CG{l_a}{m_{l_a}}{1/2}{m_s}{j_a}{1/2} \CG{l_c}{m_{l_c}}{1/2}{m_s}{j_c}{1/2} \CG{l_a}{-m_{l_a}}{l_c}{m_{l_c}}{l}{0} \\
        \notag
        & = (-1)^{-l_a+l_c} \sqrt{2l+1} \threej{l_a}{0}{l_c}{0}{l}{0} \sqrt{\frac{(2l_a+1)(2l_c+1)}{4\pi}} \sqrt{(2j_a+1)(2j_c+1)} \\
        \label{eq:angular integral specific e}
        & \quad\times \sum_{m_s,m_{l_a},m_{l_c}} (-1)^{-1/2+m_s} \threej{l_a}{m_{l_a}}{1/2}{m_s}{j_a}{-1/2} \threej{l_c}{m_{l_c}}{1/2}{m_s}{j_c}{-1/2} \threej{l_a}{-m_{l_a}}{l_c}{m_{l_c}}{l}{0} \\ 
        \notag
        & = (-1)^{-1/2-l_a+l_c} \sqrt{2l+1} \threej{l_a}{0}{l_c}{0}{l}{0} \sqrt{\frac{(2l_a+1)(2l_c+1)}{4\pi}} \sqrt{(2j_a+1)(2j_c+1)} \\
        \label{eq:angular integral specific f}
        & \quad \times \sum_{m_s,m_{l_a},m_{l_c}} (-1)^{-l_a-l_c-3/2+3m_s}(-1)^{m_s}(-1)^{l_a+l_c+1/2+m_{l_a}+m_{l_c}-m_s} \threej{j_a}{-1/2}{l_a}{m_{l_a}}{1/2}{m_s} \threej{l_c}{-m_{l_c}}{j_c}{1/2}{1/2}{-m_s} \threej{l_c}{m_{l_c}}{l_a}{-m_{l_a}}{l}{0} \\ 
        \label{eq:angular integral specific g}
        & = \sqrt{\frac{2l+1}{4\pi}} \sqrt{(2j_a+1)(2j_c+1)} \threej{j_a}{-1/2}{j_c}{1/2}{l}{0} \sqrt{(2l_a+1)(2l_b+1)} \threej{j_a}{-1/2}{j_c}{1/2}{l}{0} \sixj{j_a}{j_c}{l}{l_c}{l_a}{1/2} \\ 
        \label{eq:angular integral specific h}
        & = -\sqrt{\frac{2l+1}{4\pi}} \sqrt{(2j_a+1)(2j_c+1)} \threej{j_a}{-1/2}{j_c}{1/2}{l}{0} \threej{j_a}{1/2}{j_c}{-1/2}{l}{0} \, \Pi(l_a+l_c+l) \\ 
        \label{eq:angular integral specific i}
        & =-\sqrt{\frac{2l+1}{4\pi}} \sqrt{(2j_a+1)(2j_c+1)} \threej{j_a}{-1/2}{l}{0}{j_c}{1/2} \threej{j_a}{-1/2}{j_c}{1/2}{l}{0} \, \Pi(l_a+l_c+l) \, .
    \end{align}
\end{subequations}
In step~\eqref{eq:angular integral specific b} we expanded the spinor spherical harmonics using Eq.~\eqref{eq:bi-spinor}. In~\eqref{eq:angular integral specific c} we used $\left( Y_l^m \right)^* = (-1)^m \, Y_l^{-m}$ and the orthonormality of $\chi_{1/2,m_s}$. In~\eqref{eq:angular integral specific d} we introduced $m_{l_{a(c)}} \coloneq 1/2-m_{s_{a(c)}}$ and, because the $m$ quantum numbers on the right side of the Clebsch-Gordan coefficients must equal the one on the left side to have a non-vanishing result, we summed over it, then we summed over $m_{s_a}$, and renamed $m_s \coloneq m_{s_c}$. Finally we solved the integral of three spherical harmonics $\int Y_{l_1}^{m_1}Y_{l_2}^{m_2}Y_{l_3}^{m_3} = \sqrt{\frac{(2l_1+1)(2l_2+1)}{4\pi (2l_3+1)}}(-1)^{m_3}\CG{l_1}{m_1}{l_2}{m_2}{l_3}{-m_3}\CG{l_1}{0}{l_2}{0}{l_3}{0}$. In~\eqref{eq:angular integral specific e} we switched Clebsch-Gordan coefficients into 3-$j$ symbols using~\eqref{eq:CG-3j relation}. In~\eqref{eq:angular integral specific f} we multiplied by powers of $(-1)$ with overall product equal to 1, making manifest the possibility to use~\eqref{eq:3j sum to 6j}, in next step. In~\eqref{eq:angular integral specific h} we used~\eqref{eq:3j * 6j}. Finally, to get to the last step we used identities~\eqref{eq:3j properties}.\\
Comparing~\eqref{eq:angular integral specific i} with~\eqref{eq:WE specific} we find
\begin{equation}
    \label{eq:reduced matrix element}
    \langle j_a,l_a\|Y_l\|j_c,l_c\rangle = (-1)^{-j_a-1/2} \sqrt{\frac{2l+1}{4\pi}} \sqrt{(2j_a+1)(2j_b+1)} \threej{j_a}{-1/2}{j_c}{1/2}{l}{0} \Pi(l_a+l_c+l) \, ,
\end{equation}
leading to the following form for the general matrix element
\begin{equation}
    \label{eq:angular integral general}
    \bra{j_a,l_a,m_a}Y_l^m\ket{j_c,l_c,m_c} = (-1)^{m_a+1/2} \sqrt{\frac{2l+1}{4\pi}} \sqrt{(2j_a+1)(2j_c+1)} \threej{j_a}{-m_a}{l}{m}{j_c}{m_c} \threej{j_a}{-1/2}{j_c}{1/2}{l}{0} \, \Pi(l_a+l_c+l) \, .
\end{equation}
Using this latter formula in Eq.~\eqref{eq:one-scalar exchange stretched}, and noticing that $\Pi(l_1+l_2+l) = \Pi(l_1'+l_2'+l)$ because of Eq.~\eqref{eq:l'}, we finally find
\begin{align}
\notag
    &\bra{a,b}\Delta(\omega)\ket{c,d} = \frac{y_e^2}{4\pi} \omega_m \sum_l\left[ \sum_m \threej{j_a}{-m_a}{l}{m}{j_c}{m_c}\threej{j_b}{-m_b}{l}{-m}{j_d}{m_d} \right] \\
\notag
    & \quad \times (-1)^{m_a+m_b}(2l+1)\sqrt{(2j_a+1)(2j_b+1)(2j_c+1)(2j_d+1)} \threej{j_a}{1/2}{j_c}{-1/2}{l}{0} \threej{j_b}{1/2}{j_d}{-1/2}{l}{0}\Pi(l_a+l_c+l)\Pi(l_b+l_d+l) \\ 
    & \quad \times \int dx \, dy \, R_l(\omega_m,r_>,r_<) \left[ G_{n_a,\kappa_a}G_{n_c,\kappa_c}-F_{n_a,\kappa_a}F_{n_c,\kappa_c} \right](x) \left[ G_{n_b,\kappa_b}G_{n_d,\kappa_d}-F_{n_b,\kappa_b}F_{n_d,\kappa_d} \right](y)\,.
\end{align}

\newpage

\section{The imaginary part of the boson exchange matrix element}
\label{app:Cutkosky}
\begin{figure}
    \begin{subfigure}{0.33\linewidth}
        \begin{fmffile}{Energy-shift}
            \begin{fmfgraph*}(70,30)
            \fmfstraight
                \fmfleft{i1,i2}
                \fmfright{o1,o2}
                \fmffreeze
                \fmf{dbl_plain}{i1,v11,o1}
                \fmf{dbl_plain}{i2,v21,o2}
                \fmffreeze
                \fmf{dashes}{v21,v11}
                \fmfv{decor.shape=circle,decor.size=.15cm}{v11,v21}
                \fmflabel{a}{i1}
                \fmflabel{b}{i2}
                \fmflabel{b}{o1}
                \fmflabel{a}{o2}
            \end{fmfgraph*}
        \end{fmffile}\\[1em]
        \caption{}
        \label{fig:one-scalar energy no exchange}
    \end{subfigure}
    \hspace{4em}
    \begin{subfigure}{0.33\linewidth}
        \begin{fmffile}{Energy-SE}
            \begin{fmfgraph*}(120,30)
            \fmfstraight
                \fmfleft{i1}
                \fmfright{o1}
                \fmffreeze
                \fmf{dbl_plain}{i1,v1,v2,v3,o1}
                \fmffreeze
                \fmf{dashes,left}{v1,v3}
                \fmflabel{a}{i1}
                \fmflabel{a}{o1}
                \fmfv{decor.shape=circle,decor.size=.15cm}{v1}
                \fmfv{decor.shape=circle,decor.size=.15cm}{v3}
            \end{fmfgraph*}
        \end{fmffile}\\[1em]
        \caption{}
        \label{fig:SE energy}
    \end{subfigure}\caption{\justifying
    Feynman diagrams contributing to the energy shift due to one-scalar exchange (on the left) and SE (on the right).}
\end{figure}
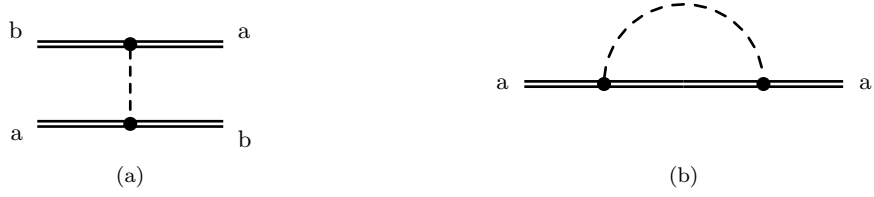

From Eq.~\eqref{eq:Rl} we see that, for $m_\phi < \omega$, the matrix element $\bra{a,b}\Delta(\omega)\ket{c,d}$ contains both a real and an imaginary part. We are interested in processes involving the exchange of scalar particles between electrons, for which the matrix elements are
\begin{subequations}
\label{eqs:matrix element one-scalar}
    \begin{gather}
    \label{eq:matrix element one-scalar no exchange}
         \bra{a,b}\Delta(0)\ket{a,b} \, ,\\
    \label{eq:matrix element one-scalar exchange}
         - \bra{a,b}\Delta(\varepsilon_a-\varepsilon_b)\ket{b,a} \, ,
    \end{gather}
\end{subequations}
where the minus sign in Eq.~\eqref{eq:matrix element one-scalar exchange} derives from the switch of two fermions.
In such matrix elements, the real part is the contribution to the energy shift, whereas the imaginary part corresponds to the decay rate. In the case of a lithium-like ion, the electron in the state $2s$ is in the ground state, implying that no decay can occur. Therefore, the imaginary part of such matrix elements should cancel in total. Only Eq.~\eqref{eq:matrix element one-scalar exchange} gives a complex results, as for Eq.~\eqref{eq:matrix element one-scalar no exchange} each value of the scalar mass is bigger than $\omega = 0$. The whole imaginary contribution of one-scalar exchange comes then from Fig.~\ref{fig:one-scalar energy no exchange}.
Studying the one-loop SE diagram of Fig.~\ref{fig:SE energy}, it can be shown (see Ref.~\cite{Shabaev2002}) that its contribution reads
\begin{equation}
\label{eq:SE energy full}
    \Delta E_\text{SE} = \frac{i}{2\pi} \int d\omega \sum_n^{\varepsilon_n\neq\varepsilon_a} \frac{\bra{a,n}\Delta(\omega)\ket{n,a}}{\varepsilon_a-\omega-\varepsilon_n(1-i0)} \, ,
\end{equation}
where $n$ denotes all the possible states, with the exception of those that are degenerate with $a$. If among all the allowed states there exists $b$ as well, then that particular piece of the sum is
\begin{equation}
\label{eq:SE energy n=b}
    \frac{i}{2\pi} \int d\omega \frac{\bra{a,b}\Delta(\omega)\ket{b,a}}{\varepsilon_a-\omega-\varepsilon_b(1-i0)} = \bra{a,b}\Delta(\varepsilon_a-\varepsilon_b)\ket{b,a} \, .
\end{equation}
The latter expression has the opposite sign of Eq.~\eqref{eq:matrix element one-scalar exchange}. 
Note that this occurs for lithium-like ions with $a=2s$ and $b=1s$, but not for helium-like ions, where $a=b=1s$. In the latter case, there are no contributions of the form~\eqref{eq:matrix element one-scalar exchange} since $a=b$, and therefore no imaginary part arises; consistently, the SE energy for the state $a=1s$ does not admit another intermediate state $1s$. The helium-like case is thus coherent, as both electrons occupy the ground states, having no decay channel available. 

We have seen that the imaginary part is completely canceled by the SE. One might then argue that, together with the imaginary part, the real part -- and therefore the whole matrix element~\eqref{eq:matrix element one-scalar exchange} -- should also cancel. However this does not lead to an inconsistency, since the real one-scalar exchange and the SE contributions are computed separately, rather than evaluating only the SE while omitting intermediate states corresponding to one-scalar exchange.

\end{document}